\documentclass{amsart}
\usepackage{eurosym}
\usepackage{amsmath, amssymb}
\usepackage{graphicx}
\usepackage{color}
\usepackage{textcomp}

\usepackage[USenglish]{babel}
\usepackage[T1]{fontenc}
\usepackage[applemac]{inputenc}
\usepackage[nodayofweek,level]{datetime}
\usepackage{lmodern}
\usepackage{tikz}
\usepackage{graphicx}
\usepackage{amsmath}
\usepackage{amssymb}
\usepackage{amsthm}
\usepackage{color}
\newcommand{\argmin}{\operatornamewithlimits{argmin}}
\usepackage{hyperref}
\usepackage{float}
\usepackage{graphicx} 
\graphicspath{{figures/}}

\DeclareSymbolFont{calletters}{OMS}{cmsy}{m}{n}
\DeclareSymbolFontAlphabet{\mathcal}{calletters}

\def\be{\begin{eqnarray}}
\def\ee{\end{eqnarray}}

\def\b*{\begin{eqnarray*}}
\def\e*{\end{eqnarray*}}

\def \E{\mathbb{E}}

\def \P{{\mathbb P}}

\def \R{\mathbb{R}}

\def \eps{\varepsilon}

\def\Cc{{\cal C}}

\def\Fc{{\cal F}}
\def\Gc{{\cal G}}

\def\Kc{{\cal K}}

\def\Nc{{\cal N}}

\def\Pc{{\cal P}}

\def\Uc{{\cal U}}

\def\Xc{{\cal X}}

\def \Prod{\displaystyle\prod}

\def\no{\noindent}
\def\x{\times}

\def\={\;=\;}
\def\.{\;.}

\def\eps{\varepsilon}

\def \1{{\bf 1}}
\def \ep{\hbox{ }\hfill{ ${\cal t}$~\hspace{-5.1mm}~${\cal u}$   } }

\def \proof{{\noindent \bf Proof. }}
\def \ep{\hbox{ }\hfill$\Box$}

 \def\normeL2#1{\left\|{#1}\right\|_{L^2}}

 \newtheorem{thm}{Theorem}[section]
 
 \newtheorem{lem}[thm]{Lemma}
 
 \theoremstyle{definition}
 \newtheorem{defn}[thm]{Definition}
 \theoremstyle{remark}
 \newtheorem{rem}[thm]{Remark}
 
 \newcommand{\PP}{\mathbb{P}}
 
 \newcommand{\X}{\mathcal{X}}

\newcommand{\bea}{\begin{eqnarray}}
\newcommand{\bes}{\begin{subequations}}
\newcommand{\ees}{\end{subequations}}
\newcommand{\bgt}{\begin{gather}}
\newcommand{\egt}{\begin{gather}}
\newcommand{\eea}{\end{eqnarray}}
\newcommand{\beaa}{\begin{eqnarray*}}
\newcommand{\eeaa}{\end{eqnarray*}}

\newcommand{\HH}{{\mathbb H}}

\newcommand{\EE}{{\mathbb E}}
\newcommand{\RR}{{\mathbb R}}
\newcommand{\cal}{\mathcal}

\newcommand{\LL}{{\mathrm L}}

\makeatletter
\g@addto@macro{\endabstract}{\@setabstract}
\newcommand{\authorfootnotes}{\renewcommand\thefootnote{\@fnsymbol\c@footnote}}%
\makeatother

\begin{document}



\begin{center}
  \LARGE 
  Building arbitrage-free implied volatility: \\ Sinkhorn's algorithm and variants \par \bigskip

  \normalsize
  \authorfootnotes
  Hadrien De March\footnote{hadrien.de-march@polytechnique.org}\textsuperscript{1}\textsuperscript{2} and Pierre Henry-Labord\`ere\footnote{pierre.henry-labordere@sgcib.com}\textsuperscript{3} \par \bigskip

  \textsuperscript{1}CMAP, \'Ecole Polytechnique \par
  \textsuperscript{2}Qantev\par
  \textsuperscript{3}Soci\'et\'e G\'en\'erale, Global Market Quantitative Research\par \bigskip

  
  \formatdate{01}{01}{2023}
\end{center}

\begin{abstract} We consider the classical problem of building an arbitrage-free  implied volatility surface from bid-ask quotes. We design a fast numerical procedure, for which we prove the convergence, based on  the Sinkhorn algorithm that has been recently used to solve efficiently (martingale) optimal transport problems.
\end{abstract}

\section{Introduction}

\no Building arbitrage-free  implied volatility surfaces from bid-ask quotes is a long-standing  issue. In particular, this is needed for market-makers in equity Vanillas. This is also needed for pricing exotic options when using risk-neutral models calibrated to Vanillas, as for the local volatility model \cite{dupire1997pricing} or for local stochastic volatility models \cite{lipton2002masterclass}. In this purpose, various approaches have been  considered. We review in the next section some of them and highlight their main drawbacks. A good method should be able to:
\begin{enumerate}
\item produce calendar/butterfly arbitrage-free surfaces.
\item fit market quotes perfectly within  bid/ask spreads.
\item fit smiles before earnings (with Mexican hat-shape curves).
\item fit quickly.
\end{enumerate}

\subsection{Review of literature}
\no In the following, for ease of notations, we assume  zero rates/dividends (see however remark (\ref{div}) for explanations how to include exactly  cash/yield dividends (and deterministic rates) in this framework). For completeness, we recall that the market price of a call option ${\cal C}(T,K) \in [(S_0-K)_+, S_0]$ with maturity $T$ and strike $K$ is quoted in terms of an implied volatility $\sigma_\mathrm{BS}(T,K)$ defined as the constant volatility $\sigma_\mathrm{BS}(T,K):=\sigma$ such that ${\cal C}(T,K)=\mathrm{BS}(S_0,K,\sigma \sqrt{T})$ where $S_0$ is the spot price value at $t=0$ and $\mathrm{BS}$ denotes the Black-Scholes formula:
\beaa \mathrm{BS}(S_0,K,\omega):=S_0 N(d_+)-K N(d_-). \eeaa 
Here $ d_\pm=\frac{\ln\left(\frac{ S_0}{K}\right)}{ \omega}\pm \frac{\omega}{2} $ and $N(x)$ is the standard normal cumulative distribution function.  As $\mathrm{BS} \in [(S_0-K)_+, S_0]$ is strictly increasing in $\omega$, the implied volatility is unique.

\subsubsection{SVM-based parameterization}
We consider the implied volatility associated to a stochastic volatility model (in short SVM), depending on some parameters: initial volatility, spot-volatility correlation, volatility-of-volatility, etc....  For example, one can consider an  SVM, defined by an homogeneous It\^o diffusion:
\beaa dS_t=C(S_t) a_t dW_t, \quad
da_t&=&(\cdots)dt + \sigma(a_t) dZ_t, \quad d \langle Z, W \rangle_t = \rho dt. \eeaa
As coming from a risk-neural model (i.e., $S_\cdot$ is a (local) martingale -- see \cite{henry2008analysis} for  sufficient and necessary conditions on the coefficients  of the diffusion with $C(s):=s$ for imposing that $S$ is not only a local martingale but a true martingale), the resulting implied volatility $,\sigma_\mathrm{BS }(\cdot,\cdot)$, for which $\EE[(S_T-K)_+]=\mathrm{BS}(S_0,K,\sigma_\mathrm{BS }(T,K) \sqrt{T})$, is arbitrage-free. In practice, the implied volatility can not be derived in closed-form and therefore the calibration of the parameters of the SVM on market prices can be quite time-consuming. In order to speed up this optimization, one can rely on the approximation of the  implied volatility in the short-maturity regime. At the first-order in the maturity $T$, one can derive a generic formula \cite{henry2008analysis}, obtained by using short-time asymptotics of the heat kernel on Cartan-Hadamard manifolds, for which the cut-locus is empty:
\bea \sigma_\mathrm{BS}(T,K) &\underset{T \rightarrow 0}{\sim}& \frac{ \ln\frac{K }{ S_0} }{ \int_{S_0}^K \frac{dx }{ a^*(x)} }\left( 1+ a_1(K)T + O(T^2) \right),  \nonumber \\ \quad a^*(x)&:=&\mathrm{argmin}_a d_\mathrm{geo}(x,a|S_0,a_0), \label{approximate} \eea
\no where the geodesic distance $d_\mathrm{geo}$ is
\beaa d_\mathrm{geo}(y_2,x_2|y_1,x_1):=\int_{y_1}^{y_2} \frac{F(y') }{ \sqrt{F(y')-C^2}} dy',
\eeaa
with $C$ defined by the equation $x_2-x_1=\int_{y_1}^{y_2}  \frac{C }{ \sqrt{F(y')-C^2}} dy'$, and $F(y):=\frac{2 }{ a(y)^2(1-\rho^2)}$,
with the new coordinates $x:=\int_{S_0}^S \frac{dz }{C(z)}-\rho \int_{a_0}^a \frac{u }{\sigma(u)} du$ and $y:=\sqrt{1- \rho^2} \int_{a_0}^a \frac{u }{\sigma(u)} du$.  The lengthly expression for $a_1(K)$ is not reported and can be found in \cite{henry2008analysis}. As an example, one can cite the SABR parameterization for which $C(S):=S^\beta$ with $\beta \in [0,1)$ and $a_t$ is a log-normal process. The resulting manifold is the $2d$ hyperbolic space $\HH^2$.  Let us remark that similar formulas can be also derived using large deviations (see \cite{friz2015large} for extensive references).

\no By construction, the implied volatility is  arbitrage-free  in strike as the parametrization comes from a risk-neutral model. However, the maturity $T$ should be ``small'' in order to preserve the validity of our approximate formula (\ref{approximate}). The arbitrability in maturity is not ensured as the calibration is performed by considering separately each time slice. Moreover, as our formula depends on a finite number of parameters, it is not possible to match exactly market prices. From a numerical point of view, the calibration involves a non-convex optimization, which is not guaranteed to converge. This solution only solves (4) and partially (1).

\subsubsection{Parametric form}   Another approach is to start directly with a parametrization of the implied volatility. As an example, commonly used by practitioners, we have the SVI parametrization \cite{gatheral2014arbitrage}
\beaa \sigma_{BS}(T,K) =a +b \left( \rho(k-m)+\sqrt{(k-m)^2 +\sigma^2}  \right),\eeaa \no depending on five parameters $a,b,\rho,m$ and $\sigma$. Note that this parametrization can be linked with the large maturity limit of the implied volatility in the Heston model. Despite its simplicity, the arbitrage-freeness in strike and maturity is not guaranteed, see however  \cite{gatheral2011convergence} for some conditions on the term-structures of the parameters (in maturity) which ensure an arbitrage-free surface \cite{gatheral2014arbitrage}. These limitations restrict the space of admissible parameters and  therefore this solution only solves (4) and partially (1).

\subsubsection{Discrete local volatility} One approach to impose the arbitrage-freeness in strike and maturity is to start (again) with a non-homogenous risk-neutral model. One can use a discrete local volatility \cite{andreasen2013expanded}.  Given a time grid of expiries $0:=t_0<t_1<\cdots<t_n$, call prices $c(t_i,\cdot)$ at time $t_ {i+1}$ are then taken to be solutions of the ODE:
\beaa
\left[1-\frac{1 }{2}\Delta t_i \sigma_i(k)^2 \partial_k^2\right] c(t_{i+1},k)=c(t_i,k), \quad c(0,k)=(S_0-k)_+. \label{forwardPDE}
\eeaa
\no By using for $\sigma_i(\cdot)$ a piecewise constant function, we can try to match market prices of call options. As pointed in \cite{lipton2011filling}, ``this method uses a fully implicit finite-difference scheme to compute the probability density of the underlying, stepping forward in time
and calibrating model parameters by a least-squares algorithm. Since the size
of time step is determined by market quotes, it cannot be reduced arbitrarily, so
that, while very instructive, this method clearly has limited accuracy''.  For example, with this algorithm, we were not able   to calibrate equity Vanillas exhibiting a Mexican hat form (see Figure \ref{google}),  just before earning dates. Some improvements  have been considered in \cite{lipton2011filling}.

\subsubsection{Sinkhorn algorithm} This algorithm \cite{sinkhorn1967concerning} has been popularized recently for solving quickly optimal transportation problems by \cite{cuturi2013sinkhorn}. This algorithm consists in solving an optimal transport problem by including an entropy term in order to make it strictly convex, and then the dual of this entropic optimal transport problem is solved by doing alternatively projections on the marginal distributions of the two measures transported on each other. It has been a quite hot research topic lately, (see for example \cite{peyre2017computational}, \cite{merigot2011multiscale}, or \cite{schmitzer2016stabilized} for amazing practical approaches). A Sinkhorn's algorithm including the martingale constraint was introduced by \cite{guo2017computational} in one dimension, and \cite{de2018entropic} in multi-dimensions with practical approaches. In these works, a third projection on the martingale constraint is introduced and allows to quickly solve the martingale optimal transport problem.

\subsection{Contents} In this paper, we will build a solution satisfying (1-2-3-4) by construction. The conditions (1-2-3) are automatically (and exactly) satisfied as we construct a non-parametric density fitting the Vanillas. Our approach is close in spirit to the ``Weighted Monte-Carlo approach'' based on an entropic penalisation as introduced in \cite{avellaneda2001weighted}. However, our approach use a non-degeneracy hypothesis in order to prove the existence of smooth fitting probability. We introduce the framework and the goal in Section \ref{sect:intro} to help the reader accommodate with the concepts. Then in Section \ref{sect:m-markov}, we prove the appropriate theoretical results to show the shape of the fitting model we build, and then we provide an algorithm to obtain it in practice. The convergence of our algorithm is then proved (see Theorem \ref{thm:convergence}) with a fast decay rate and therefore our numerical scheme solves (4). We finally in Subsection \ref{subsect:all_maturities} show under which condition we can extend the construction from two times to a higher number of times. We conclude with numerous examples of fitting to Equity Vanillas for various stocks and indices in Section \ref{sect:numerical}. Then Section \ref{sect:proofs} is dedicated to the proofs of the technical results.

\section{Axiomatics: Formulation}\label{sect:intro}

\no Prices of call options for different maturities $t_1 < \cdots < t_n$ and different strikes are quoted on the market. We denote by
$\Cc_i^K$ the market prices of maturity $t_i$ and strike $K \in {\cal K}_i$. The set ${\cal K}_i$ corresponds to the strikes  $K_i^1 < \cdots < K_i^{n_i}$. We denote $\Pc(\Xc)$ the probability measures on a set $\Xc$. Building an arbitrage-free implied volatility is equivalent to finding a martingale probability measure $\PP^*\in\Pc(\RR_+^n)$ that matches (exactly) this  market prices: $\PP^*$ should belong to the convex set
\beaa
{\cal M}_n=\left\{ \PP \; : \;  \EE^\PP[(S_{t_i}-K)_+]=\Cc_i^K, \quad \forall K \in {\cal K}_i, \; \EE^\PP[S_{t_i}|S_0,\cdots, S_{t_{i-1}}]=S_{t_{i-1}}, \quad i=1, \cdots ,n \right\}.
\eeaa
\no For use below, we set ${\cal C}_i^j:=\Cc_i^{K_i^j}$ and define the prices of Vertical Spreads (VS), Calendar Vertical Spreads (CVS), and Calendar Butterfly Spreads (CBS):
\beaa
\mathrm{VS}_i^j &:=& \frac{ {\cal C}_i^{j-1}-{\cal C}_i^{j} }{K_i^j -K^{j-1}_i } \quad 1 \leq j \leq n_i, \\
\mathrm{VS}_i^0&:=&1, \\
\mathrm{CVS}_ {i_1,i_2}^{j_1,j_2}&:=& C_{i_2}^{j_2}- C_{i_1}^{j_1}, \\
\mathrm{CBS}_ {i,i_1,i_2}^{j,j_1,j_2}&:=&\frac{CVS_ {i_1,i}^{j_1,j}
}{K_{i_1}^{j_1} -K_{i}^j }-\frac{ CVS_ {i,i_2}^{j,j_2}
}{K_{i}^j -K_{i_2}^{j_2}  }.
\eeaa
\no For completeness, we cite the following result that gives necessary and sufficient conditions for  arbitrage-freeness:
\begin{lem}[see \cite{carr2005note,cousot2007conditions} for proofs]\label{lemma:Mn-non-empty} ${\cal M}_n$ is non-empty if and only if for all $i=1,\cdots,n$

\no {\bf (1)}
\beaa
&&{\cal C}_i^j \geq 0,\quad 0 \leq j \leq n_i, \\
&&\mathrm{VS}_i^j \in [0,1],\quad  1 \leq j \leq n_i, \\
&&\mathrm{VS}_i^j>0 \quad\; \mathrm{if} \quad \forall \; 1 \leq j \leq n_i, \mbox{ we have } C^{j-1}_i>0.
\eeaa

\no {\bf (2)}  $\forall i_1,i_2 \in [1,n]$ s.t. $i_1<i_2$, $\forall j_1 \in [0,n_{i_1}]$, $\forall j_2 \in [0,n_{i_2}]$
\beaa
&&\mathrm{CVS}_{i_1,i_2}^{j_1,j_2} \geq 0, \quad \mathrm{if} \quad K_{i_1}^{j_1} \geq K_{i_2}^{j_2}, \\
&&\mathrm{CVS}_{i_1,i_2}^{j_1,j_2} > 0, \quad \mathrm{if} \; K_{i_1}^{j_1} > K_{i_2}^{j_2}  \; \mathrm{and} \; {\cal C}^{j_1}_{i_1}>0.
\eeaa

\no {\bf (3)} $\forall i,i_1,i_2 \in [1,n]$ s.t. $i \leq i_1$ and $i \leq i_2$, $\forall j \in [0,n_i]$, $\forall j_1 \in [0,n_{i_1}]$ , $\forall j_2 \in [0,n_{i_2}]$ s.t.  $
K_{i_1}^{j_1}<K_i^j < K_{i_2}^{j_2}$:
\beaa \mathrm{CBS}^{j,j_1,j_2}_{i,i_1,i_2} \geq 0. \eeaa

\end{lem}

\subsection*{Markovian solutions}
\no As a simplification, we could assume that $\PP^*$ should satisfy a  Markov  property and therefore belongs instead to the subset of  ${\cal M}_n$:
\beaa {\cal M}^\mathrm{Markov}_n=\{\PP\in\Pc^\mathrm{Markov} (\R^n_+)\,:\,  \EE^\PP[(S_{t_i}-K)_+]=\Cc_i^K, \; \forall K \in {\cal K}_i,  \;
\EE^\PP[S_{t_i}|S_{t_{i-1}}]=S_{t_{i-1}}, \; i=1, \cdots ,n \}. \eeaa

\no Where $\Pc^\mathrm{Markov}(\R^n_+)$ is the set of probability measures of $\Pc(\R^n_+)$ that satisfy the Markov property.

\begin{lem} ${\cal M}^\mathrm{Markov}_n$ is non-empty if and only ${\cal M}_n$ is non-empty. In particular if the market data $( {\cal C}_i)_{1 \leq i \leq n}$ are arbitrage-free, they can be attained by a martingale measure in ${\cal M}^\mathrm{Markov}_n$.
\end{lem}
\proof
\no

\no {\bf $\Longrightarrow$}: obvious.

\no {\bf $\Longleftarrow$} Take $\PP \in {\cal M}_n$.  Then by disintegration, define the marginals $\PP^{i-1}$ and $\PP^{i}$, which are in the convex order. From Strassen theorem \cite{strassen1965existence}, we can build a martingale measure  $\PP^{i-1,i}$ with marginals $\PP^{i-1}$ and $\PP^{i}$ (see e.g. \cite{henry2013explicit}, or \cite{kallblad2017optimal} for an explicit construction).  By gluing these measures, we get an element in
${\cal M}^\mathrm{Markov}_n$.
\ep

\subsection{Sequential construction}\label{subsect:sequential_build}

\no From the Markov property, an element $\PP \in {\cal M}^\mathrm{Markov}_n$ could be written as
\beaa \PP(ds_1, \cdots,ds_n)=\PP^{0,1}(ds_1) \prod_{i=2}^n \PP^{i-1,i}(ds_i|s_{i-1}), \eeaa
where the probability $\PP^{0,1} $ and $(\PP^{i-1,i})_{i=1,\cdots,n}$ are constructed as follows:

\no {\bf (1)} We choose a $\PP^{0,1}  \in {\cal M}^\mathrm{Markov}_1$ with
\beaa {\cal M}^\mathrm{Markov}_1:=\{\PP\in\Pc (\{S_0\}\x\R_+) \; &:& \;  \EE^\PP[(S_{t_1}-K)_+]=\Cc_1^K, \quad \forall K \in {\cal K}_1,  \quad
\EE^\PP[S_{t_1}|S_0]=S_{0} \}. \eeaa
\no {\bf (2)} We choose a $\PP^{1,2} \in  {\cal M}^\mathrm{Markov}_{1,2}$ with \beaa {\cal M}^\mathrm{Markov}_{1,2}(\PP^{1}):=\{\PP\in\Pc (\R^2_+) \;:\; S_{t_{1}} \overset{\PP}{\sim} \PP^{0,1}, \quad \EE^\PP[(S_{t_2}-K)_+]=\Cc_2^K, \quad
\EE^\PP[S_{t_2}|S_{t_{1}}]=S_{t_{1}} \}. \eeaa From $\PP^{1,2} \in  {\cal M}^\mathrm{Markov}_{1,2}$, we define $\PP^2$ as
 \beaa \PP^2(ds_2)=\int \PP^{1,2}(ds_1,ds_2). \eeaa
 \no {\bf (3)}  We iterate step {\bf (2)} to obtain  $(\PP^{i-1,i})_{i=3,\cdots,n}$.
 
 \begin{rem}
 Notice that the probability measures in ${\cal M}^\mathrm{Markov}_1$ and $ {\cal M}^\mathrm{Markov}_{1,2}(\P^1)$ are trivially Markov, as they are on two times.
 \end{rem}

 \subsection{Adding bid-ask prices} In practice, market prices are quoted with bid-ask prices.  Our discussion can be generalized to this case by replacing $ {\cal M}^\mathrm{Markov}_1$, ${\cal M}^\mathrm{Markov}_{1,2}$, and ${\cal M}^\mathrm{Markov}_n$ by
\beaa
\widetilde{\cal M}^\mathrm{Markov}_1&:=&\{\PP\in\Pc(\{S_0\}\x\R_+) \, : \,   \Cc^{K,\mathrm{bid}}_1 \leq \EE^\PP[(S_{t_1}-K)_+] \leq \Cc^{K,\mathrm{ask}}_1, \;\\
&&\forall K \in {\cal K}_1,  \;
\EE^\PP[S_{t_1}|S_0]=S_{0} \}.
\eeaa 
\beaa 
\widetilde{\cal M}^\mathrm{Markov}_{1,2}(\PP^{0,1})&:=&\{\PP \, : \, S_{t_1} \overset{\PP}{\sim} \PP^{0,1},\; \EE^\PP[S_{t_2}|S_{t_1}]=S_{t_1}, \; \Cc^{K,\mathrm{bid}}_2 \leq \EE^\PP[(S_{t_{2}}-K)_+] \leq \Cc^{K,\mathrm{ask}}_2, \;\\
&&\forall K \in {\cal K}_2  \}.
\eeaa 
\beaa 
\widetilde{\cal M}^\mathrm{Markov}_n&:=&\{\PP\in\Pc^\mathrm{Markov} (\R^n_+)\,:\,  \Cc^{K,\mathrm{bid}}_i \leq \EE^\PP[(S_{t_{i}}-K)_+] \leq \Cc^{K,\mathrm{ask}}_i, \; \forall K \in {\cal K}_i,  \;\\
&&\EE^\PP[S_{t_i}|S_{t_{i-1}}]=S_{t_{i-1}}, \; i=1, \cdots ,n \}.
\eeaa

\no We consider this setup in the next sections. The arbitrage-free conditions, which ensure that  $\widetilde{\cal M}^\mathrm{Markov}_{1,2}(\PP^{0,1})$ is non-empty, are given in \cite{cousot2007conditions}: we can take the same conditions than in Lemma \ref{lemma:Mn-non-empty} with redefining
\b*
\mathrm{VS}_i^j &:=& \frac{ {\cal C}_i^{K_{j-1},\mathrm{bid}}-{\cal C}_i^{K_j,\mathrm{ask}} }{K_i^j -K^{j-1}_i } \quad 1 \leq j \leq n_i, \\
\mathrm{CVS}_ {i_1,i_2}^{j_1,j_2}&:=& C_{i_2}^{K_{j_2},\mathrm{ask}}- C_{i_1}^{K_{j_1},\mathrm{bid}}, \\
\mathrm{CBS}_ {i,i_1,i_2}^{j,j_1,j_2}&:=&\frac{C_{i}^{K_{j},\mathrm{ask}}- C_{i_1}^{K_{j_1},\mathrm{bid}}
}{K_{i_1}^{j_1} -K_{i}^j }-\frac{ C_{i_2}^{K_{j_2},\mathrm{bid}}- C_{i}^{K_{j},\mathrm{ask}}
}{K_{i}^j -K_{i_2}^{j_2}  }.
\e*

\begin{rem}[Cash/yield dividends] \label{div} \no We assume here that the spot process $S_t$ jumps down by the dividend amounts $D_i(S_{\tau_i^-}) = \beta_i \,S_{\tau_i^-}+\alpha_i$, paid at the dates $0 < \tau_1 < \dots \tau_n < T$, and that between dividend dates it follows a diffusion. By setting $S_t=A(t)+B(t)X_t$ (see \cite{henry2009calibration} for formulas for $A$  and $B$ as  functions of $(\alpha_i,\beta_i)$), one obtains that $X_t$ is a martingale.  Call options on $S$ can therefore be written as call options on $X$. One can then apply our construction to $X$  and deduce then call options on $S$. Using this mapping, we will assume no dividends/zero rates in the following.
\end{rem}

\section{Building an element in $\widetilde{\cal M}^\mathrm{Markov}_n$}\label{sect:m-markov}

\subsection{Existence of the solution between two times}

\no As explained in Subsection \ref{subsect:sequential_build}, the goal is to first build a coupling $\P^{i-1,i}$ between each consecutive maturities of the call options.

\subsubsection{Notation}

For this purpose we introduce generic notation: let $S_1$ and $S_2$ be two real random variables. Let $0<K_1<...<K_k$, $\Kc:=\{K_i:1\le i \le k\}$, $\left(\Cc^{K,{\rm bid/ask}}\right)_{K\in\Kc}\in \R^{2k}$, and $\P^1$ a probability distribution on $\R$ with a finite support ${\rm supp}(\P^1)$, where for any probability $\P$ we denote by ${\rm supp}(\P)$ its support. The goal is to build a coupling in
\beaa 
\widetilde{\cal M}\left(\P^1,\left(\Cc^{K,{\rm bid/ask}}\right)_{K\in\Kc}\right)&:=&\left\{\PP \, : \, S_1 \overset{\PP}{\sim} \PP^1,\; \EE^\PP[S_{2}|S_{1}]=S_{1},\right.\\
&&\Cc^{K,\mathrm{bid}} \leq \EE^\PP[(S_{2}-K)_+] \leq \Cc^{K,\mathrm{ask}}, \;\left. \forall K \in {\cal K}\right\}
\eeaa 
For the sake of simplicity, in the rest of the paper, we denote $\widetilde{\cal M}:=\widetilde{\cal M}\left(\P^1,\left(\Cc^{K,{\rm bid/ask}}\right)_{K\in\Kc}\right)$.

\subsubsection{Optimisation problem approach} An element $\PP^* \in \widetilde{\cal M}$ can be obtained by minimizing a convex lower semi-continuous functional ${\Fc}$:
\bea \label{opt1}
\inf_{\PP \in \widetilde{\cal M}} {\Fc}(\PP)={\Fc}(\PP^*), \quad \PP^* \in \widetilde{\cal M}.
\eea
Then an optimisation of the dual problem associated will allow to obtain explicitly this optimizer, hence allowing to obtain an element in $\widetilde{\cal M}$.

\subsubsection{Choice of $\Fc_1$} Let $m_0$ be a prior measure on $\R^2$. We choose $(\omega_K)_{K \in {\cal K}_1} \in \RR^{n_1}_+$ and consider the regularized Kullback-Leibler functional:
\beaa
&&
{\Fc}(\PP):=  \EE^{\PP}\left[  \ln \frac{d\PP }{d{m_0}}-1\right] {+\sum_{K \in \Kc} \frac{1 }{2 \omega_K}\big(\EE^\PP[(S_1-K)_+]-\Cc^{K,\mathrm{mid}}_1\big)^2 },
\eeaa
where we denote $\Cc^{K,\mathrm{mid}}:= \frac{\Cc^{K,\mathrm{bid}}+\Cc^{K,\mathrm{ask}}}{2}$. This equation depends on a prior measure ${m_0}$ on $\RR_+$, left unspecified for the moment. Notice that by introducing dual variables $v_K \in \RR$, for each $K \in {\cal K}$, therefore ${\Fc}$ may also be written as
\beaa
{\Fc}
(\PP):=\EE^{\PP}\left[ \ln \frac{d\PP }{d{m_0}}-1\right] -{\inf_{v \in \RR^{{\cal K}}} \sum_{K \in {\cal K}} v_K\left(\Cc^{K,\mathrm{mid}}- \EE^\PP[(S_1-K)_+]\right)+ \frac{1 }{2}v_K^2 \omega_K }.
\eeaa

\no Notice that we choose the form $\EE^{\PP}\left[ \ln \frac{d\PP }{d{m_0}}-1\right]$ over $\EE^{\PP}\left[ \ln \frac{d\PP }{d{m_0}}\right]$ because it gives exactly the same solutions, but with simpler formulas.

\subsubsection{Existence theorem}

\no The following condition will guarantee the existence of solutions for \ref{opt1}.

\begin{defn}\label{def:non-degen}
We say that $\left(m_0, \P^1, \left(\Cc^{K,{\rm bid/ask}}\right)_{K\in\Kc}\right)$ is non-degenerate if up to denoting $K_0:=0$ and $K_{k+1}:=\infty$, and setting $\Cc^{0,{\rm bid}} := \Cc^{0,{\rm ask}} := \E^{\P}[S_1]$ and $\Cc^{\infty,{\rm bid}} := \Cc^{\infty,{\rm ask}} := 1$, we may find $\Cc\in \R^{k+2}$ such that for all $0\le l \le k+1$, we have

\no\rm{(i)} $\Cc^{K_l,{\rm bid}}\le \Cc_l\le \Cc^{K_l,{\rm ask}}$,

\no\rm{(ii)} $( M_{\rm call}^{-1}\Cc)_l>0$,

\no\rm{(iii)} $\Cc_l>\E^{\P^1}[(S_1-K_l)_+]$, if $1\le l\le k$,

\no\rm{(iv)} the projection of $m_0$ on $S_1$ has support ${\rm supp}(\P^1)$, which is a finite support.

\no\rm{(v)} $m_0[\{s\}\x (K_l, K_{l+1})]>0$, for $s\in{\rm supp}(\P^1)$ if $l\le k$,

\no\rm{(vi)} $m_0[\{s\}\x (-\infty, s)]>0$ and $m_0[\{s\}\x (s, \infty)]>0$, for $s\in{\rm supp}(\P^1)$,

\no where $ M_{\rm call}:=\big((K_{l_{2}}-K_{l_1})_+\big)_{0\le l_1,l_2\le k+1}$, with the convention $(K_{k+1}-K_l)_+ := (K_l-K_{k+1})_+ := 1$ for all $l$.
\end{defn}

\begin{rem}
Notice that {\rm (ii)} in Definition \ref{def:non-degen} is equivalent to a kind of no-arbitrage condition: it is equivalent to the fact that if we define a non-negative payoff $f(s):=\sum_{i=0}^{k+1}\lambda_i(s-K_i)_+\ge 0$, then we have $\sum_{i=0}^{k+1}\lambda_i\Cc_i\ge 0$, with equality if and only if $\lambda_0 = ...= \lambda_{k+1} = 0$.
\end{rem}

\begin{thm}\label{thm:minimG_elem_M}
We assume that $\left(m_0, \P^1, \left(\Cc^{K,{\rm bid/ask}}\right)_{K\in\Kc}\right)$ is non-degenerate.

\no Then the minimization (\ref{opt1}) is attained by $\PP^* \in \widetilde{\cal M}$ with
\beaa
\PP^*(ds_1,ds_2)={m_0}(ds_1,ds_2) e^{- \sum_{K \in {\cal K}} V_K^* (s_{2}-K)_+-u^*(s_1) -h^*(s_1)(s_2-s_1)  },
\eeaa
where $u^*$, $h^*$, and $  V^*$ solve the strictly convex unconstrained minimization:
 \beaa
 \inf_{V \in \RR^{{\cal K}}, u, h \in \LL^1(\PP^1)}&& \Gc(u,h,V),
 \eeaa
 where
\beaa
\Gc(u,h,V) &:=& \EE^{\PP^1}[u(S_1)]+\sum_{K \in {\cal K}}f^{K,\mathrm{bid}/\mathrm{ask}}(V_K,\omega_K)+\sum_{K \in {\cal K}}  V_K \Cc^{K,\mathrm{mid}} \\
&&+\EE^{{m_0}}\left[  e^{- \sum_{K \in {\cal K}} V_K (S_{2}-K)_+ -u(S_1)-h(S_1)(S_2-S_1)  }\right].
\eeaa

\beaa
\mbox{and}\quad f^{K,\mathrm{bid}/\mathrm{ask}}(V,\omega):=&\frac{V^2 \omega }{2}, &\mbox{if }  \Delta \Cc^{K,\mathrm{bid}}\leq V \omega \leq  \Delta \Cc^{K,\mathrm{ask}} \\
 :=&\Delta \Cc^{K,\mathrm{ask}} V -\frac{(\Delta \Cc^{K,\mathrm{ask}})^2 }{2 \omega}
 , &\mbox{if }  \Delta \Cc^{K,\mathrm{ask}}< V \omega \\
 :=& \Delta \Cc^{K,\mathrm{bid}} V -\frac{(\Delta \Cc^{K,\mathrm{bid}})^2 }{2 \omega}
 , &\mbox{if }   \Delta \Cc^{K,\mathrm{bid}}> V \omega.
\eeaa
Here $ \Delta {\cal C}^{\mathrm{bid}/\mathrm{ask}}:= {\cal C}^{\mathrm{bid}/\mathrm{ask}}-{\cal C}^\mathrm{mid}$. 
\end{thm}

\no The proof of Theorem \ref{thm:minimG_elem_M} is reported to Section \ref{sect:proofs}.

\subsubsection{Dependence on the prior}

\no We consider two prior densities $\PP_0$ and $\PP_0'$. By definition, the vanillas constructed using the two priors satisfy the equations for all $K \in {\cal K}_1$:
\beaa \Cc^{K,\mathrm{mid}}_1+\partial_Vf(V_K,\omega_K)- {\cal C}^\mathrm{model}(K,\PP_0)=0 \\
\Cc^{K,\mathrm{mid}}_1+\partial_V f(V'_K,\omega_K)- {\cal C}^\mathrm{model}(K,\PP_0')=0
\eeaa By taking the difference, we get
\beaa
|{\cal C}^\mathrm{model}(K,\PP'_0)-{\cal C}^\mathrm{model}(K,\PP_0)|&=&|\partial_V f(V'_K,\omega_K)-\partial_V f(V_K,\omega_K)| \\
&\leq& \omega_K |V'_K-V_K|.
\eeaa

\subsection{Sinkhorn's algorithm to find the solution between two times}

\subsubsection{Solutions to partial optimisation of $\Gc$}

\no Let $s_1\in {\rm supp}(\P^1)$, the zero of the gradient with respect to $u$ is given by the equation:
\bea
e^{-u(s_1)}=\frac{\PP^1(s_1) }{I_u(h(s_1),V(\cdot),s_1)} \label{u1} \eea
where we have set
\beaa
I_u(\theta,V,s_1):=\int e^{- \sum_{K \in {\cal K}} V_K (s_{2}-K)_+-\theta(s_2-s_1)} m_0(s_1,ds_2) = e^{u(s_1)}\partial_{u(s_1)}\Gc(u,h,V),  \label{I_u}
\eeaa

\no The zero of the gradient with respect to $h(s_1)$ is given by the equation:  $h(s_1):=\theta$ is the unique zero of
\bea
I_h( \theta,V,s_1):=0, \label{h}
\eea
where
\beaa
I_h(\theta,V,s_1):=\int e^{- \sum_{K \in {\cal K}} V_K (s_{2}-K)_+ -\theta(s_2-s_1)} (s_2-s_1){m_0}(s_1,ds_2) = e^{u(s_1)}\partial_{h(s_1)}\Gc(u,h,V).   \label{I_h}
\eeaa
In practice, this may be done thanks to a 1D Newton algorithm on the function $h(s_1)\mapsto \min_{h(s_1)}\Gc(u,h,V)$, see Subsections 3.3.3 and 3.3.5 in \cite{de2018entropic}.

\no For use below, we also introduce  for all $Q \in {\cal K}$:
\beaa I_Q(h,V,s_1):=\int (s_2-Q)_+ e^{- \sum_{K \in {\cal K}} V_K (s_{2}-K)_+ -h(s_2-s_1)} {m_0}(s_1,ds_2)  = e^{u(s_1)}\partial_{V_Q}\Gc(u,h,V).
\eeaa

\subsubsection{Sinkhorn's algorithm in a nutshell}

\begin{enumerate}
\item[I] Start with $h:=0$, $u:=0$ and $V_K:=0$ for all $K \in {\cal K}$.
\item[II] Projection on $(u,h)$: Solve equations \eqref{u1} and \eqref{h} for all $s_1\in{\rm supp}(\P^1)$.
\item[III] Solve the strictly convex smooth finite-dimensional unconstrained minimization over $V$:
\beaa  \inf_{V_K \in \RR}&&\Gc(u,h,V),
\eeaa
with
\b*
\Gc(u,h,V)&:=& \E^{\P^1}[u]+ \sum_{K \in {\cal K}}f_2^{K,\mathrm{bid}/\mathrm{ask}}(V_K,\omega_K)+\sum_{K \in {\cal K}}  V_K \Cc^{K,\mathrm{mid}}\\
&&+\EE^{{m_0}}\left[  e^{- \sum_{K \in {\cal K}} V_K (S_{2}-K)_+ -h(S_1)(S_2-S_1)-u(S_1)  }\right].\label{eq:minsinkhorn}
\e*

\no Notice that we provide a practical approach for this step in Subsection \ref{subsect:prior}.

\item[IV] Iterate steps (II)-(III) until convergence.
\end{enumerate}

\subsubsection{Convergence}

%

\no Notice that as $u,h\in \LL^1(\PP^1)$ can be identified with vectors in $\R^{|{\rm supp}(\P^1)|}$ as $\P^1$ has finite support. We will abuse notation and do this confusion in the rest of the paper.

\begin{thm}[Convergence rate]\label{thm:convergence}
The map $\Gc$ reaches a minimum $\Gc^*$ at some $x^*\in \R^{2|{\rm supp}(\P^1)|+|\Kc|}$ if and only if $\left(\P^1, \left(\Cc^{K,{\rm bid/ask}}\right)_{K\in\Kc}\right)$ is non-degenerate.

\no In this case, let $x_0 = (u_0,h_0,V_0)\in \R^{2|{\rm supp}(\P^1)|+|\Kc|}$, and for $n\ge 0$, let the $n^{th}$ iteration of the well-defined martingale Sinkhorn algorithm:
\b*
x_{n+1/2} &:=& \left(u_n,h_n,V_{n+1}:=\argmin_\psi \Gc(u_n,h_n,\cdot)\right),\\
x_{n+1} &:=& \left(u_{n+1}:=\argmin_u \Gc(\cdot,\cdot,V_{n+1}),h_{n+1}:=\argmin_h \Gc(\cdot,\cdot,V_{n+1}),V_{n+1}\right).
\e*
Then $\P_n(ds_1,ds_2):={m_0}(ds_1,ds_2) e^{- \sum_{K \in {\cal K}} V_K^n (s_{2}-K)_+-u^n(s_1) -h^n(s_1)(s_2-s_1)  }$ is a martingale probability with marginal $\P^1$ on $S_1$ and we may find $0<\lambda<1$, and $M>0$ such that
\b*
\Gc(x_n)-\Gc^*\le
\lambda^{n}\big(\Gc(x_0)-\Gc^*\big)&\mbox{and}& \max_{K\in\Kc}{\rm dist}_{[\Cc^{K,{\rm bid}},\Cc^{K,{\rm ask}}]}\left(\E^{\P_n}[(S_2-K_i)_+]\right)\le M\sqrt{\lambda}^n,
\e*
for all $n\ge 0$.
\end{thm}

\no The proof of Theorem \ref{thm:convergence} is reported to Section \ref{sect:proofs}.

\begin{rem}
Notice that
\b*
\nabla \Gc &=& \sum_{s_1\in{\rm supp}(\P^1)}\left(\P^1[\{s_1\}]-\P\circ (S_1)^{-1}[\{s_1\}]\right)e_{s_1}+\sum_{s_1\in{\rm supp}(\P^1)}\left(\E^\P[S_2-s_1,S_1=s_1]\right)e_{|{\rm supp}(\P^1)|+s_1}\\
&+&\sum_{i=1}^k\left(C_K-\E^\P[(S_2-K_i)_+]\right)e_{2|{\rm supp}(\P^1)|+i},
\e*
where $C_K\in[\Cc^{K,{\rm bid}},\Cc^{K,{\rm ask}}]$, and $(e_i)_{1\le i \le 2|{\rm supp}(\P^1)|+|\Kc|}$ is the canonical basis. Therefore, this gradient is a crucial estimate of the mismatch of $\P$ in terms of first marginal, martingale property, and correctness of the call prices it gives.
\end{rem}

\begin{rem}
We might obtain better stability and speed of convergence for the minimising of $\Gc$ by using an implied Newton minimization algorithm (see 3.3.5. in \cite{de2018entropic}). This algorithm consists of applying a truncated Newton algorithm on $\widetilde{\Gc}(V):= \min_{u,h}\Gc(u,h,V)$ which is strongly convex and smooth like $\Gc$, see Proposition 3.2 in \cite{de2018entropic}. This algorithm would have the same complexity, as we use a Newton algorithm of the same dimension $|\Kc|$ for the partial minimization in $V$ during phase (III) of the Sinkhorn algorithm, and the partial minimisation of $\Gc$ in $u$ and $h$ would be equivalent to steps (I) and (II). However, we can see from \cite{de2018entropic} that the convergence is much faster.
\end{rem}

\begin{rem}
Even though the criterion from Definition \ref{def:non-degen} may not be easy to compute, trying to solve the entropic minimization reveals if a solution exists as otherwise the map $\Gc$ diverges to $-\infty$. In this case there is an arbitrage between the call prices and $\P^1$.
\end{rem}

\subsection{Extension to all the maturities}\label{subsect:all_maturities}

Recall that we have maturities $0<t_1<...<t_n$, and for $1\le i\le n$, call strikes $\Kc_i$ and their bid/ask spread $(\Cc^{K,\mathrm{bid/ask}}_i)_{K\in\Kc_i}$. In order to state a result for all the maturities and build an element from $\widetilde{\cal M}^\mathrm{Markov}_n$, we need to define a new global non-degeneracy condition. We look for the solution with a prior measure
\beaa 
m_n(s_0,...,s_n):=m^{-1,0}(s_0)\Prod_{i=1}^n m^{i-1,i}(ds_i|s_{i-1})
\eeaa 
with $m^{-1,0}:= \delta_{S_0}$, and measures with finite supports $m^{i-1,i}$ for $1\le i\le n$. Also similar to for Definition \ref{def:non-degen}, we index the call strikes for convenience: $(K^i_l)_{1\le l\le k_i}:=\Kc_i$, where $k_i:=|\Kc_i|$ for $1\le i \le n$.

\begin{defn}\label{def:non-degen-all}
We say that $\left(m_n, \left(\Cc^{K,{\rm bid/ask}}_i\right)_{1\le i\le n,K\in\Kc_i}\right)$ is non-degenerate if up to denoting $K^i_0:=0$ and $K^i_{k+1}:=\infty$, and setting $\Cc^{0,{\rm bid}}_i := \Cc^{0,{\rm ask}}_i := S_0$ and $\Cc^{\infty,{\rm bid}}_i := \Cc^{\infty,{\rm ask}}_i := 1$ for $1\le i\le n$, we may find $\Cc^i\in \R^{k_i+2}$ such that for all $1\le i\le n$ and $0\le l \le k_i+1$, we have

\no\rm{(i)} $\Cc^{K^i_l,{\rm bid}}_i\le \Cc^i_l\le \Cc^{K^i_l,{\rm ask}}_i$,

\no\rm{(ii)} $( M_{\rm call}^{-1}\Cc^i)_l>0$,

\no\rm{(iii)} $\Cc^{K^i_l,\rm bid}_i>\Cc^{K^{i-1}_{l'},\rm ask}_{i-1}+(K^{i-1}_{l'}-K^i_l)_+$, for some $1\le l'\le k_{i-1}$, if $1\le i\le k$,

\no\rm{(iv)} the supports ${\rm supp}(m^{i-2, i-1})\circ S_{t_{i-1}}$ and ${\rm supp}(m^{i-1, i})\circ S_{t_{i-1}}^{-1}$ are equal and finite,

\no\rm{(v)} $m^{i-1, i}[\{s\}\x (K^i_l, K^i_{l+1})]>0$, for $s\in{\rm supp}(m^{i-1, i})\circ S_{t_{i-1}}^{-1}$ if $l\le k_i$,

\no\rm{(vi)} $m_0[\{s\}\x (-\infty, s)]>0$ and $m_0[\{s\}\x (s, \infty)]>0$, for $s\in{\rm supp}(m^{i-1, i})\circ S_{t_{i-1}}^{-1}$,

\no where $ M_{\rm call}:=\big((K^i_{l_2}-K^i_{l_1})_+\big)_{0\le l_1,l_2\le k_i+1}$, with the convention $(K^i_{k_i+1}-K^i_l)_+ := (K^i_l-K^i_{k_i+1})_+ := 1$ for all $l$.
\end{defn}

\begin{thm}
\label{thm:minimG_elem_M_all}
We assume that $\left(m_n, \left(\Cc^{K,{\rm bid/ask}}_i\right)_{1\le i\le n,K\in\Kc_i}\right)$ is non-degenerate.

\no Then we may find $\P_0,...,\P_{n-1}\in \Pc(\R_+^*)$ such that for all $1\le i \le n$, the minimization
\beaa
\inf_{\PP \in \widetilde{\cal M}\left(\P^{i-1},\left(\Cc^{K,{\rm bid/ask}}\right)_{K\in\Kc_i}\right)} {\Fc}(\PP)={\Fc}(\PP^{i-1,i}).
\eeaa

is attained by $\PP^{i-1,i} \in \widetilde{\cal M}\left(\P^{i-1},\left(\Cc^{K,{\rm bid/ask}}\right)_{K\in\Kc_i}\right)$ with
\beaa
\PP^{i-1,i}(ds_1,ds_2)={m^{i-1,i}}(ds_1,ds_2) e^{- \sum_{K \in {\cal K}_i} V_K^i (s_{i}-K)_+-u^{i-1}(s_{i-1}) -h^{i-1}(s_{i-1})(s_i-s_{i-1})  },
\eeaa
where $u^{i-1}$, $h^{i-1}$, and $  V^{i}$ solve the strictly convex unconstrained minimization:
 \beaa
 \inf_{V^i \in \RR^{{\cal K}_i}, u, h \in \LL^1(\PP_{i-1})}&& \Gc_{i-1,i}(u,h,V),
 \eeaa
 where
\beaa
\Gc_{i-1,i}(u,h,V) &:=& \EE^{\PP_{i-1}}[u(S_{t_{i-1}})]+\sum_{K \in {\cal K}_i}f_i^{K,\mathrm{bid}/\mathrm{ask}}(V_K,\omega_K)+\sum_{K \in {\cal K}_i}  V_K \Cc^{K,\mathrm{mid}}_i \\
&&+\EE^{{m^{i-1,i}}}\left[  e^{- \sum_{K \in {\cal K}_i} V_K (S_{t_{i}}-K)_+ -u(S_{t_{i-1}})-h(S_{t_{i-1}})(S_{t_{i}}-S_{t_{i-1}})  }\right].
\eeaa

\beaa
\mbox{and}\quad f_i^{K,\mathrm{bid}/\mathrm{ask}}(V,\omega):=&\frac{V^2 \omega }{2}, &\mbox{if }  \Delta \Cc_i^{K,\mathrm{bid}}\leq V \omega \leq  \Delta \Cc_i^{K,\mathrm{ask}} \\
 :=&\Delta \Cc_i^{K,\mathrm{ask}} V -\frac{(\Delta \Cc_i^{K,\mathrm{ask}})^2 }{2 \omega}
 , &\mbox{if }  \Delta \Cc_i^{K,\mathrm{ask}}< V \omega \\
 :=& \Delta \Cc_i^{K,\mathrm{bid}} V -\frac{(\Delta \Cc_i^{K,\mathrm{bid}})^2 }{2 \omega}
 , &\mbox{if }   \Delta \Cc_i^{K,\mathrm{bid}}> V \omega.
\eeaa
Here $ \Delta {\cal C}_i^{\mathrm{bid}/\mathrm{ask}}:= {\cal C}_i^{\mathrm{bid}/\mathrm{ask}}-{\cal C}_i^\mathrm{mid}$. 

Finally, we have that
\bea\label{eq:found_m_markov} 
&m_n(ds_0,...,ds_{n}) e^{- \sum_{i=0}^{n-1}\left(u(s_{i-1})+h(s_{i-1})(s_{i}-s_{i-1})+\sum_{K \in {\cal K}_{i+1}} V_K^i (s_{i+1}-K)_+ \right)  }\in \widetilde{\cal M}^\mathrm{Markov}_n.
\eea 
\end{thm}

\no The proof of Theorem \ref{thm:minimG_elem_M_all} is reported to Section \ref{sect:proofs}.

\begin{rem}
Theorem \ref{thm:minimG_elem_M_all} allows to have an algorithm to compute by recurrence all the probabilities $\P_1,...,\P_{n_1}$ starting at $\P_0:=\delta_0$, and then build the probability in $\widetilde{\cal M}^\mathrm{Markov}_n$ thanks to applying $n$ times the Sinkhorn's algorithm.
\end{rem}

\begin{rem}
Finding a model generating volatility splines that is Markovian allows to have an approach that is computationally efficient. However, by duality, it might still generate models that hold arbitrages, indeed in \eqref{eq:found_m_markov}, the hedges $h^i$ are only dependent on the latest value of the asset $S_t$ in order to be Markovian. Therefore there might be arbitrages exploiting path-dependent hedges $h(s_{t_1},s_{t_2},...,S_t)$. In order to handle them, we should consider them as argument of $\Gc$ however it makes the computational cost explode, as for one Sinkorn projection we need to take into consideration $size_grid^{n}$ values of $h^n(S_{t_1},...,S_{t_n})$.
\end{rem}

\begin{rem}
As we solve the problem building $\P^{i,i+1}$ after having built $\P^{i-1,i}$, we needed \rm{(iii)} from Definition \ref{def:non-degen-all} to prevent a situation in which the probability $\P_i$ might not fit the condition \rm{(iii)} of Definition \ref{def:non-degen} because of the call prices available in the intervals $[\Cc^{K,\rm bid},\Cc^{K,\rm ask}]$. This situation might happen in tail prices where the bid-ask spread if large like we can see on Figure \ref{google}. In this case we could solve a global entropic optimal transport problem, by projecting on all the times iteratively, like it is done in \cite{benamou2015iterative}. Then we would have a Sinkorn's algorithm with $2n$ projections instead of $2$. Another simpler solution would be to select in advance call prices in the bid-ask spreads that have no arbitrage.  We have not encountered this situation in our numerical experiments using real market data.
\end{rem}

\section{Numerical experiments}\label{sect:numerical}

\subsection{ Speed-up: Choice of a prior}\label{subsect:prior}

\no We take ${m_0}(ds_1,ds_2)=\mathbf{1}_{s_1\ge 0}\PP_{\sigma_0 }(ds_1) \PP(ds_2 |s_1)$ where $\PP_{\sigma_0 }(ds_1)$ is the discrete approximation of a normal density with volatility $\sigma_0$ and under $\PP$:
\beaa
S_2=S_1+ \sigma(S_1)\sqrt{t_2-t_1} Z, \quad Z \in \mathrm{N}(0,1), \quad
\sigma(S_1)=\sigma_0 S_1^{\beta},
\eeaa
where $\sigma_0$ and $\beta$ are two parameters. We choose $\sigma_0$ and $\beta$ by minimizing  the least-square problem:
\beaa
&&\inf_{\sigma_0,\beta} \sum_{K \in {\cal K}} \left(\EE^{{m_0}}[ (S_2-K)_+]-\Cc^{K,\mathrm{mid}} \right)^2 \\
&=&\inf_{\sigma_0,\beta} \sum_{K \in {\cal K}} \left( \EE^{\PP_{\sigma_0 }}[\mathrm{B}(S_1,t_2-t_1,K,\sigma_0 S_1^{\beta})]-\Cc^{K,\mathrm{mid}} \right)^2,
\eeaa
with
\beaa
\mathrm{B}(s,t,K,\sigma):=\frac{1}{2} (s-K) \text{erf}\left(\frac{K-s}{\sqrt{2} \sigma  \sqrt{t}}\right)+\frac{\sigma
   \sqrt{t} e^{-\frac{(K-s)^2}{2 \sigma ^2 t}}}{\sqrt{2 \pi }}.
   \eeaa

\no Notice that by the fact that $S_2$ is normally-distributed when conditioned on $S_1$, the integration over $s_2$ can be performed exactly thanks to the definition of the functions $I_u$, $I_h$ and $I_Q$, defined above. These functions' values can be written in closed-form.

\begin{rem}[Explicit formulas]\label{rmk:closed_formulas}
For completeness, we give the formulas, obtained  with Mathematica, that we use in our numerical implementation. Let $A:=\frac{K_1-s_1 }{\sigma}$, $B:=\frac{K_2-s_1 }{\sigma}$ and $\sigma:=\sigma(s_1)\sqrt{t_2-t_1}$. We have
\beaa
&&\int_{K_1}^{K_2} e^{ \alpha s_2}  \PP(ds_2|s_1)=
\frac{1}{2} e^{\frac{\alpha ^2 \sigma ^2}{2}+\alpha
   s_1} \left(\text{erf}\left(\frac{B-\alpha  \sigma
   }{\sqrt{2}}\right)-\text{erf}\left(\frac{A-\alpha  \sigma
   }{\sqrt{2}}\right)\right). \\
 &&2 \sqrt{2 \pi}\int_{K_1}^{K_2} e^{ \alpha s_2} (s_2-s_1) \PP(ds_2|s_1)= \sigma  e^{\alpha  s_1} \left(2 e^{A \alpha
   \sigma -\frac{A^2}{2}}-\sqrt{2 \pi } \alpha \sigma
   e^{\frac{\alpha ^2 \sigma ^2}{2}}
   \text{erf}\left(\frac{A-\alpha  \sigma
   }{\sqrt{2}}\right)\right. \\
   &&\left.+\sqrt{2 \pi } \alpha  \sigma
   e^{\frac{\alpha ^2 \sigma ^2}{2}}
   \text{erf}\left(\frac{B-\alpha  \sigma }{\sqrt{2}}\right) -2e^{\alpha  B \sigma -\frac{B^2}{2}}\right).\\
 && 2 \sqrt{2 \pi }\int_{K_1}^{K_2} e^{ \alpha s_2} (s_2-K) \PP(ds_2|s_1)= e^{\alpha  s_1} \left(2 \sigma  e^{A \alpha
   \sigma -\frac{A^2}{2}}-\sqrt{2 \pi } e^{\frac{\alpha ^2
   \sigma ^2}{2}} \text{erf}\left(\frac{A-\alpha  \sigma
   }{\sqrt{2}}\right) \left(\alpha  \sigma
   ^2-K+s_1\right) \right. \\
   && \left. +\sqrt{2 \pi } e^{\frac{\alpha ^2
   \sigma ^2}{2}} \text{erf}\left(\frac{B-\alpha  \sigma
   }{\sqrt{2}}\right) \left(\alpha  \sigma
   ^2-K+s_1\right)  -2 \sigma  e^{\alpha  B \sigma
   -\frac{B^2}{2}}\right). \\
 &&2 \sqrt{2 \pi }\int_{K_1}^{K_2} e^{ \alpha (s_2-s_1)} (s_2-K)(s_2-Q) \PP(ds_2|s_1)=2 \sigma  \left(e^{\alpha  A
   \sigma -\frac{A^2}{2}} (\sigma  (\alpha  \sigma +A)-K-Q+2
   s_1) \right.\\ &&\left.+e^{-\frac{1}{2} B (B-2 \alpha  \sigma )} (-\sigma
   (\alpha  \sigma +B)+K+Q-2 s_1)\right)\\
   &&+\sqrt{2 \pi }
   e^{\frac{\alpha ^2 \sigma ^2}{2}}
   \left(\text{erf}\left(\frac{B-\alpha  \sigma
   }{\sqrt{2}}\right)-\text{erf}\left(\frac{A-\alpha  \sigma
   }{\sqrt{2}}\right)\right) \left(\sigma ^2-\left(-\alpha  \sigma
   ^2+K-s_1\right) \left(\alpha  \sigma
   ^2-Q+s_1\right)\right).
   \eeaa
   \no The last formula is used for computing the hessian $\partial_V^2  {\cal G}_{12}$.
\end{rem}

\begin{rem}[Other formulas] \label{remint}
Note that we have
\beaa
\EE^{{m_0}}\left[  e^{- \sum_{K \in {\cal K}} V_K (S_{2}-K)_+ -h(S_1)(S_2-S_1)-u(S_1)  }\right]
=\EE^{\PP_{\sigma_0}}\left[ I_u(h(S_1),V(\cdot),S_1) e ^{-u(S_1)  } \right].
\eeaa and
\beaa
\EE^{\PP_{\sigma_0}}\left[   (S_{2}-Q)_+ e^{- \sum_{K \in {\cal K}} V_K (S_{2}-K)_+ -h(S_1)(S_2-S_1)-u(S_1)  }\right]
=\EE^{\PP_{\sigma_0}}\left[ I_Q(h(S_1),V(\cdot),S_1) e ^{-u(S_1)  } \right].
\eeaa

From Remark \ref{remint}, \eqref{eq:minsinkhorn} can be written exactly as
\b*
\Gc(u,h,V)&=&  \E^{\P^1}[u]+ \sum_{K \in {\cal K}}f^{K,\mathrm{bid}/\mathrm{ask}}(V_K,\omega_K)+\sum_{K \in {\cal K}}  V_K \Cc^{K,\mathrm{mid}}\\
&&+\EE^{\PP_{\sigma_0}}\left[  I_u(h(S_1),V,S_1) e ^{-u(S_1)  }\right]. \nonumber
\e*
The gradients with respect to $V_K$ can also be written  as
\b*
\partial_{V_K} f^{K,\mathrm{bid}/\mathrm{ask}}(V_K,\omega_K)+\Cc^{K,\mathrm{mid}}
-\EE^{\PP_{\sigma_0}}\left[ I_K(h(S_1),V,S_1) e ^{-u(S_1)  } \right].
\e*

And the hessians with respect to $V_{K_1}$ and $V_{K_2}$ are given by
\b*
\delta_{K_1=K_2}\partial^2_{V_{K_1}^2} f^{K_1,\mathrm{bid}/\mathrm{ask}}(V_{K_1},\omega_{K_1})+\Cc^{K,\mathrm{mid}}
-\EE^{\P}\left[ (S_2-K_1)_+(S_2-K_2)_+ \right].
\e*
\no With $\P := e^{- \sum_{K \in {\cal K}} V_K (S_{2}-K)_+ -h(S_1)(S_2-S_1)-u(S_1)  }\PP_{\sigma_0}$.

\no These formulas may be used to do the partial minimisation in $V$ with a "pure" Newton method, or with a quasi-Newton method which only requires the gradient.

 \end{rem}
 
  \subsection{Numerical examples}
\no In practice, we take $\omega_K=\Lambda|\Cc^{K,\mathrm{ask}}_1-\Cc^{K,\mathrm{bid}}_1|$ with
$\Lambda=0.1$ in our numerical examples. The minimization over $V$ is performed using a
modified Newton method and a user-supplied Hessian. In order to have easier computations thanks to the closed formulas displayed in Remark \ref{rmk:closed_formulas}, we use as a reference measure $m_0(ds_1):=\P_0(ds_1)\mathbf{1}_{s_1\ge 0}$, where $\P_0$ is the Gaussian measure $\Nc(S_0,\sigma_0^2t_1)$, properly normalized on $\RR_+$, and where $\sigma_0$ is chosen to minimize the criterion:
\beaa
&&\inf_{\sigma_0} \sum_{K \in {\cal K}_1} \left(\EE^{\PP_0}[ (S_1-K)_+]-\Cc^{K,\mathrm{mid}}_1 \right)^2 =\inf_{\sigma_0} \sum_{K \in {\cal K}_1} \left( \mathrm{B}(S_0,t_1,K,\sigma_0 )-\Cc^{K,\mathrm{mid}}_1 \right)^2,
\eeaa
with
\beaa \mathrm{B}(s,t,K,\sigma):=\frac{1}{2} (s-K) \text{erf}\left(\frac{K-s}{\sqrt{2} \sigma  \sqrt{t}}\right)+\frac{\sigma
   \sqrt{t} e^{-\frac{(K-s)^2}{2 \sigma ^2 t}}}{\sqrt{2 \pi }}.
   \eeaa

\begin{figure}[h]
\begin{center}
\includegraphics[width=7cm,height=5cm]{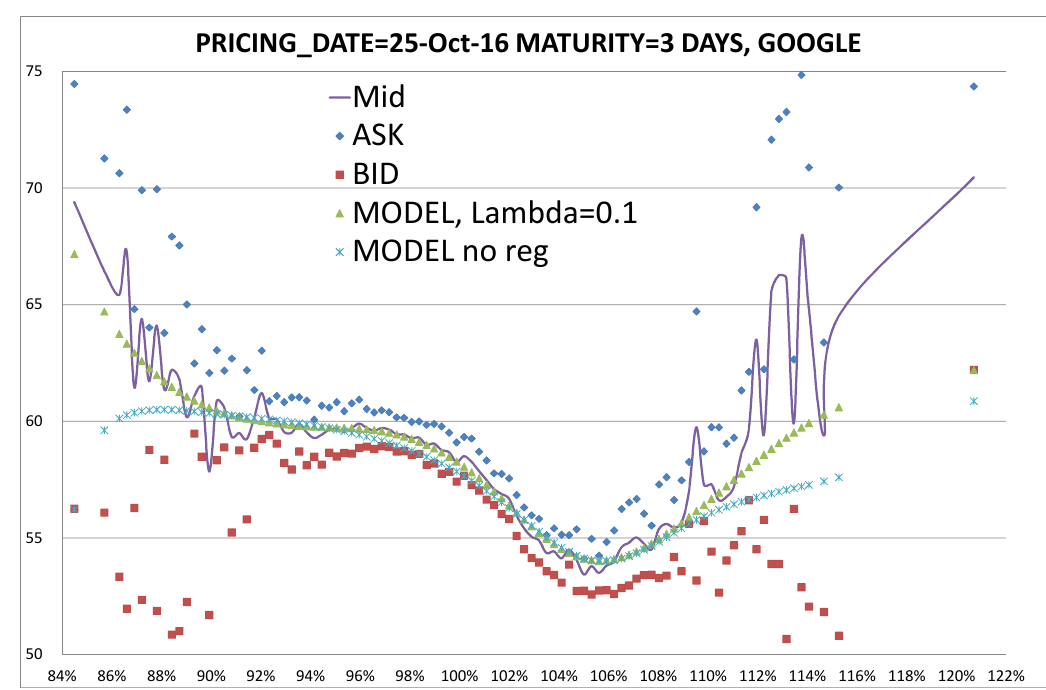}
\includegraphics[width=7cm,height=5cm]{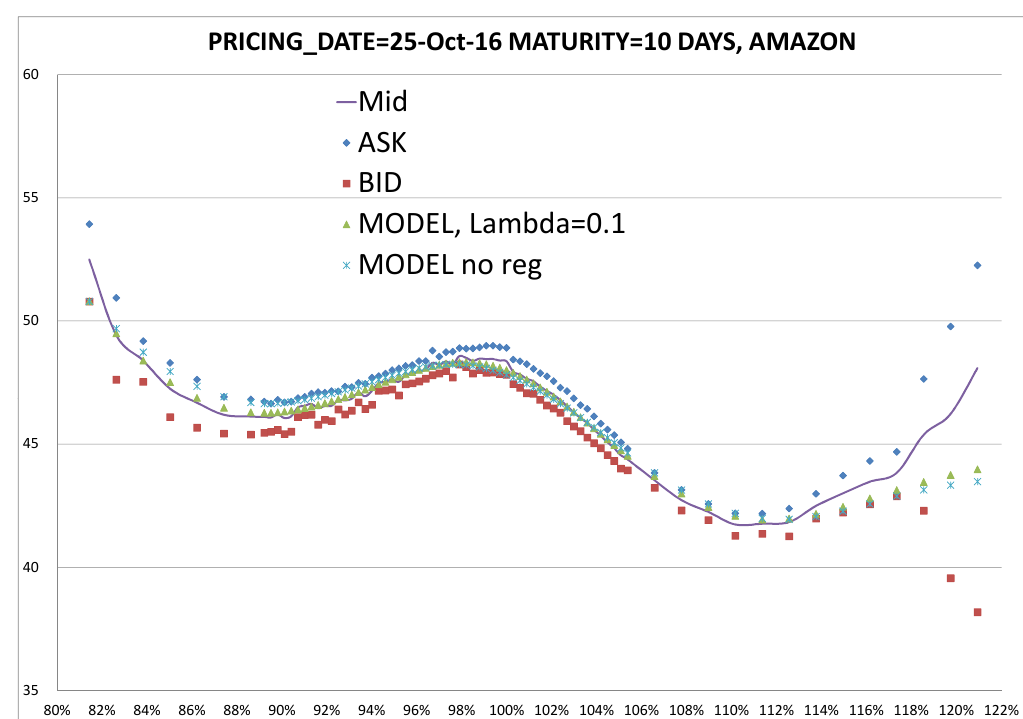}
\caption{Computational time = 0.1 s. Left: GOOGLE. Right: AMAZON. The plots denoted ``Model no reg'' mean that we have chosen $\Lambda=\infty$.}
\label{google}
\end{center}
\end{figure}

\begin{figure}[h]
\begin{center}\label{earning2}
\includegraphics[width=7cm,height=5cm]{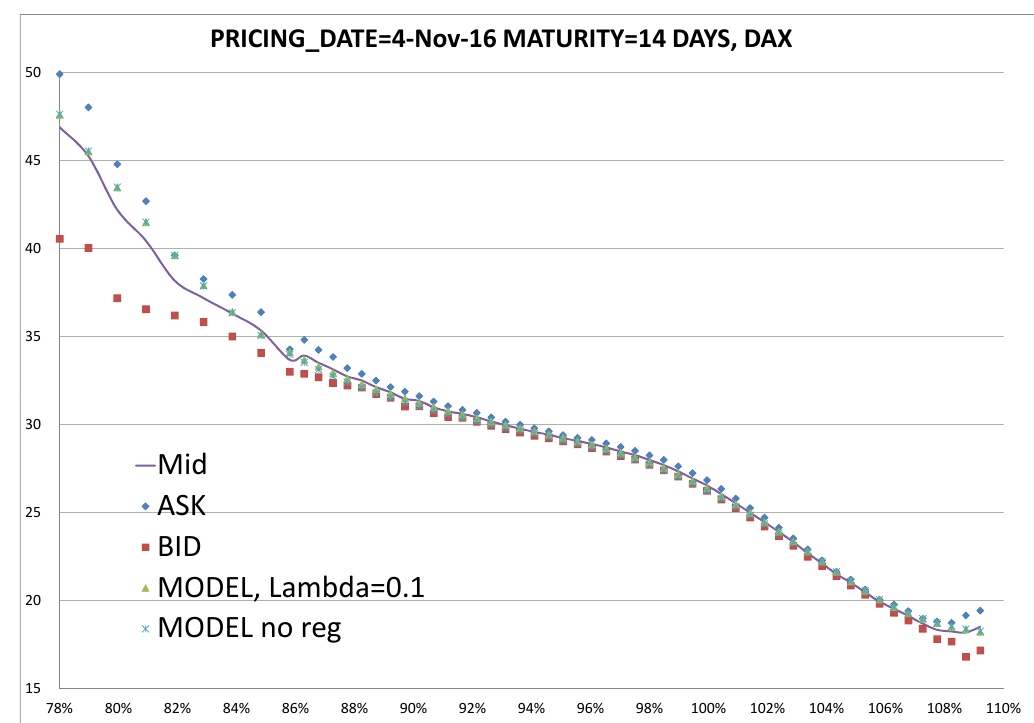}
\includegraphics[width=7cm,height=5cm]{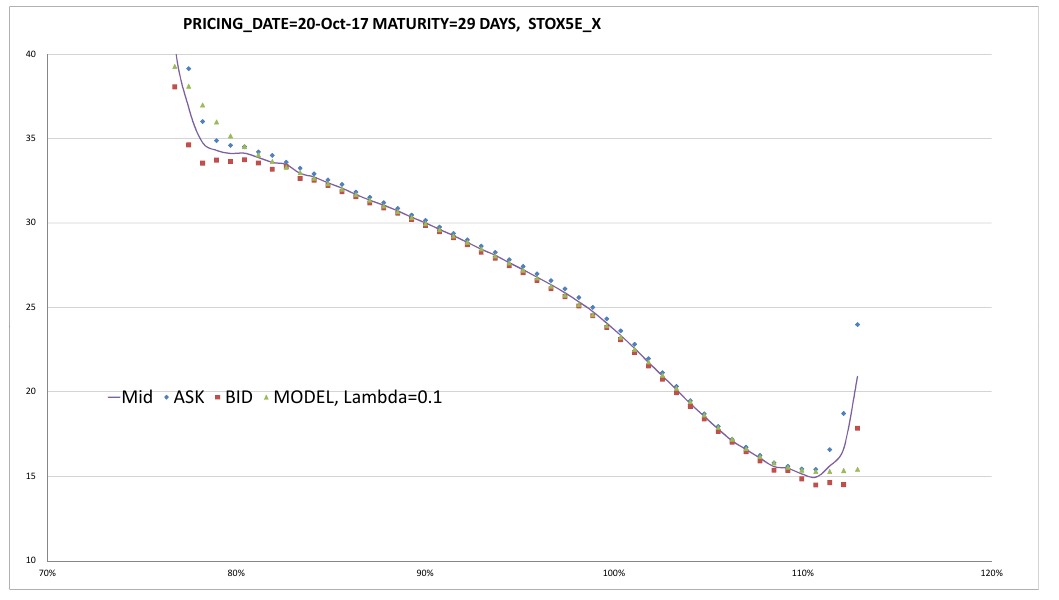}
\end{center}
\caption{Computational time = 0.1 s. Left: DAX. Right: EURO STOXX 50.}
\end{figure}

\no In Figure \ref{google}, we  show examples of calibration with two stocks (Google \& Amazon) near earnings. By construction, the fit is perfect (within the bid/ask spread) and arbitrage-free. \no In Figure \ref{earning2},  we consider two indices (Dax \& Euro Stoxx 50).

\no Below, we list some numerical examples involving numerous equity stocks/indices with various liquidity/maturities: Soci\'et\'e G\'en\'erale, Danone,  Apple, SP500.

\begin{figure}[tpb]
\begin{center}
\includegraphics[width=5cm,height=4cm]{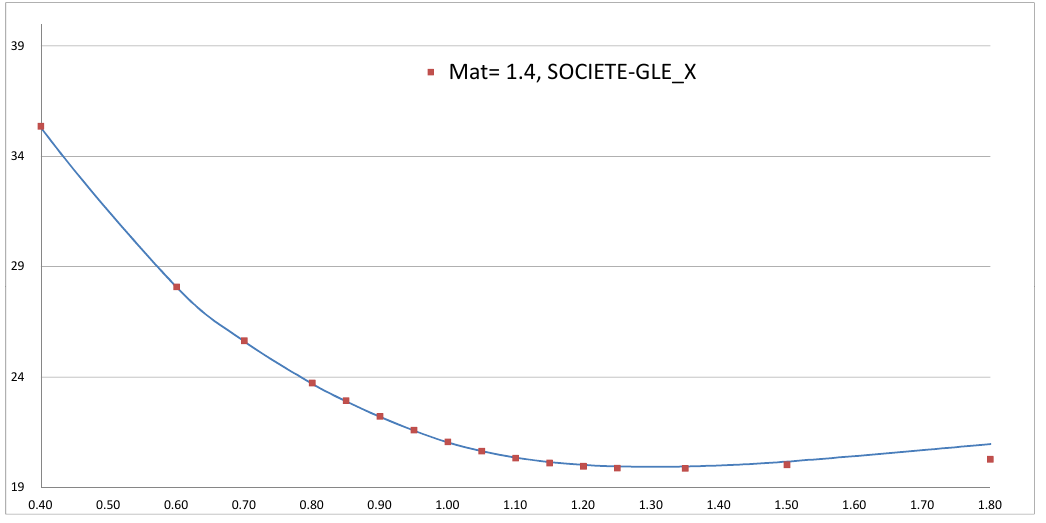}
\includegraphics[width=5cm,height=4cm]{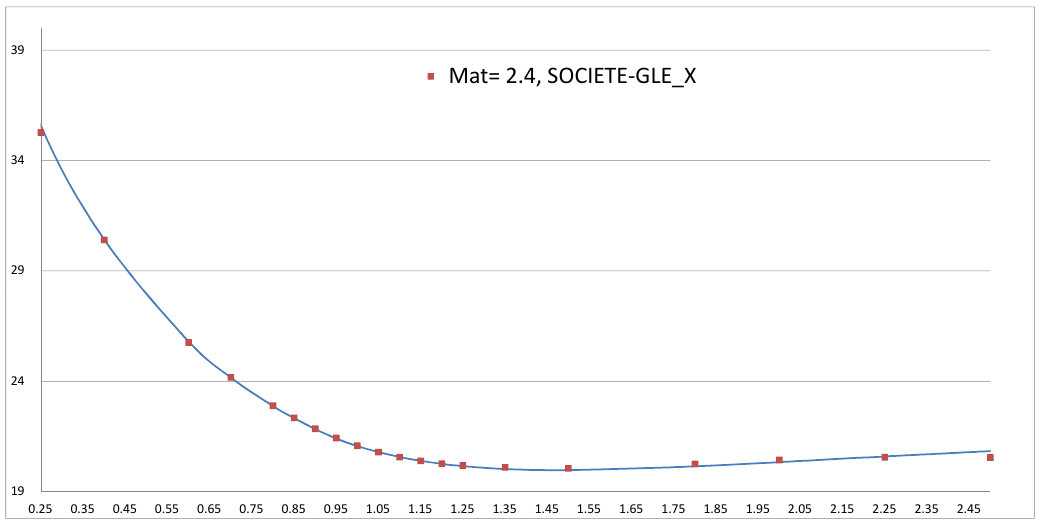}
\end{center}
\caption{SOCIETE-GENERALE.}
\end{figure}

\begin{figure}[tpb]
\begin{center}
\includegraphics[width=5cm,height=4cm]{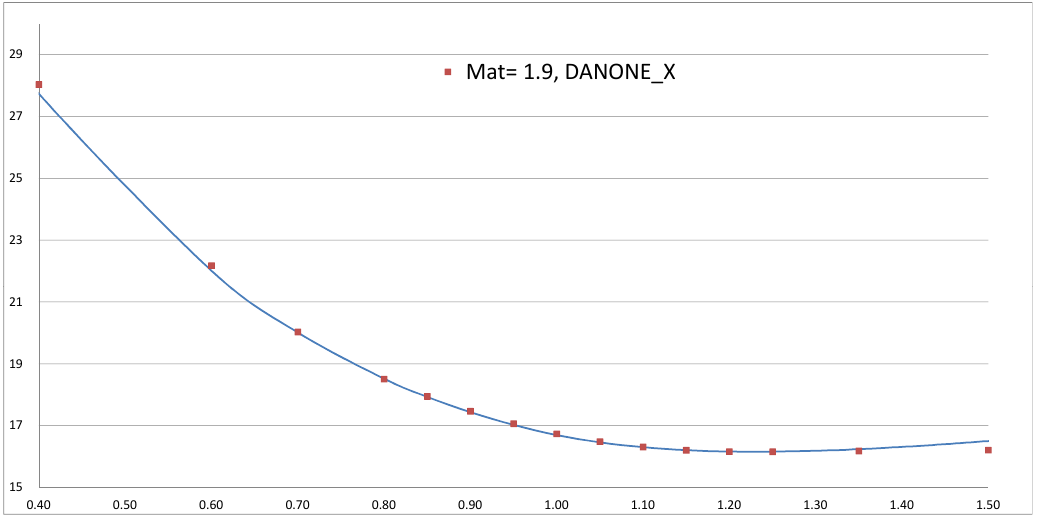}
\includegraphics[width=5cm,height=4cm]{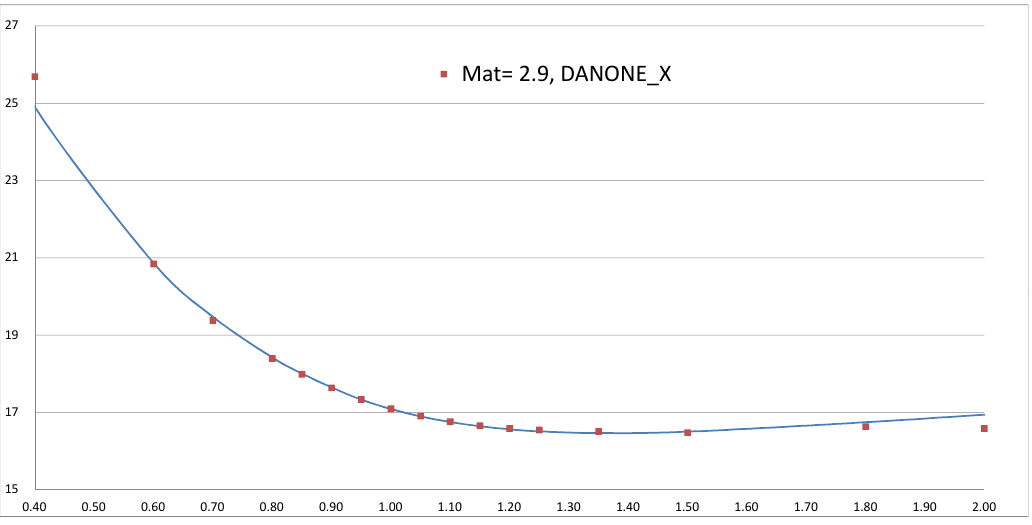}
\end{center}
\caption{DANONE.}
\end{figure}

\begin{figure}[tpb]
\begin{center}
\includegraphics[width=5cm,height=4cm]{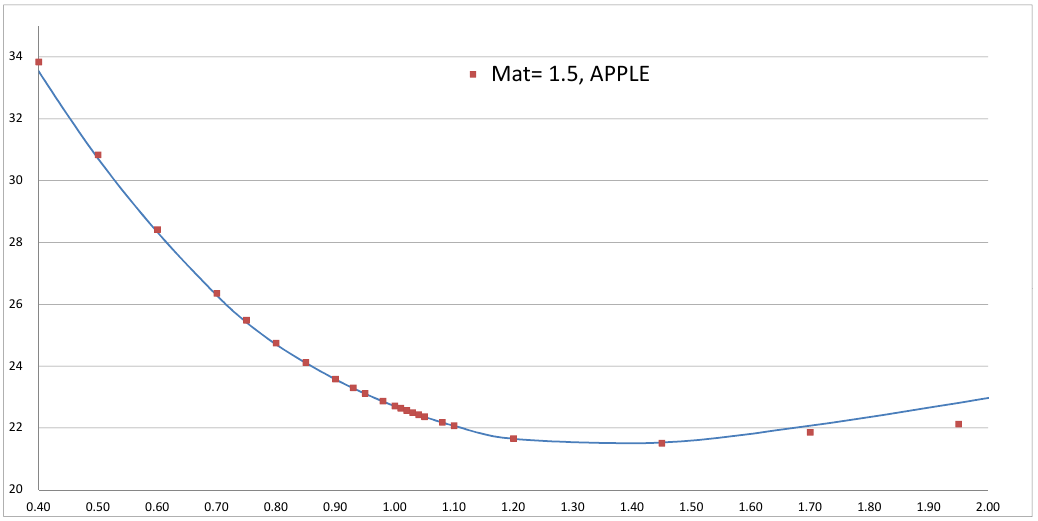}
\includegraphics[width=5cm,height=4cm]{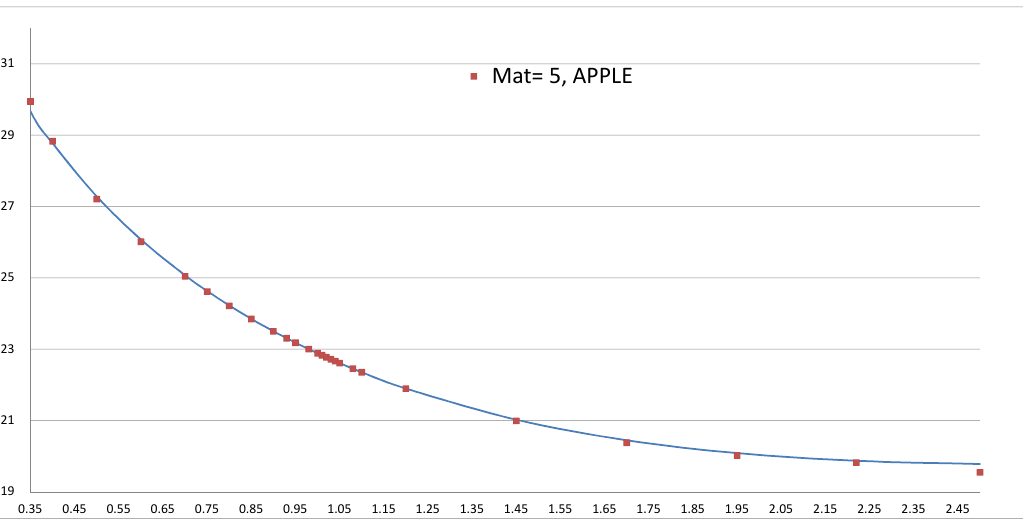}
\end{center}
\caption{APPLE.}
\end{figure}

\begin{figure}[tpb]
\begin{center}
\includegraphics[width=5cm,height=4cm]{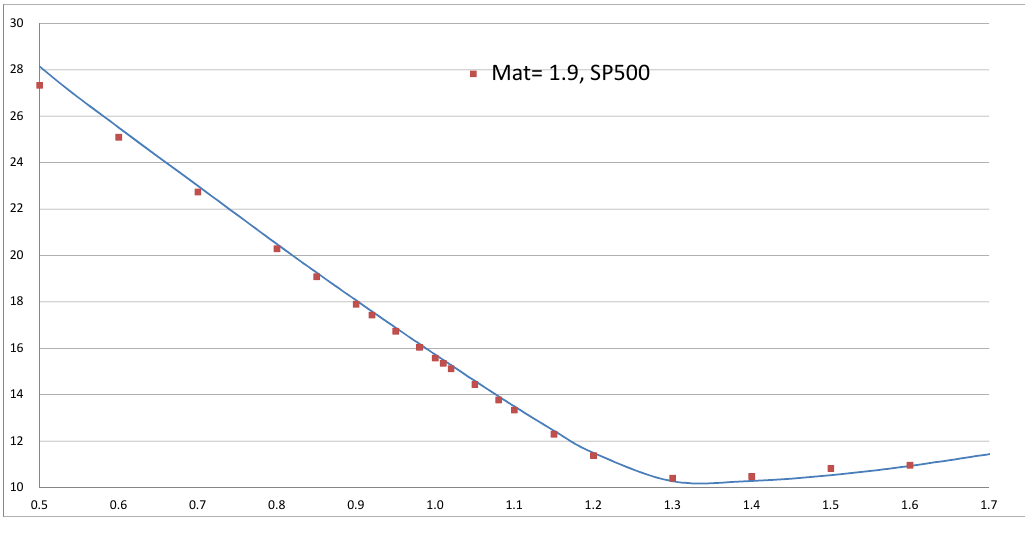}
\includegraphics[width=5cm,height=4cm]{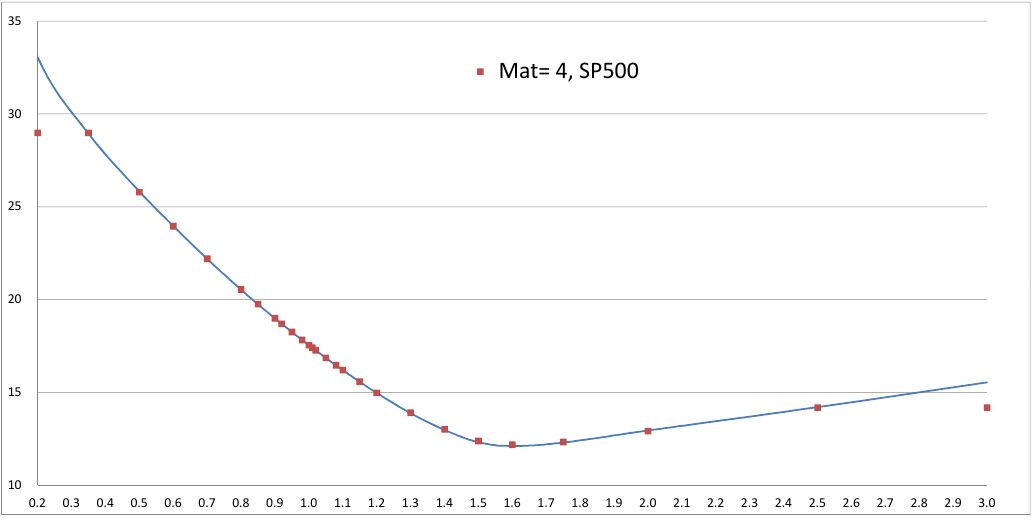}
\end{center}
\caption{SP500.}
\end{figure}

\section{Proofs of the results.}\label{sect:proofs}

\begin{lem}\label{lemma:minGinM}
Let $(u^*,h^*,V^*)$ be a minimiser of $\Gc$, then if we define
\beaa
\PP^*(ds_1,ds_2)&:=&{m_0}(ds_1,ds_2) e^{- \sum_{K \in {\cal K}} V_K^* (s_{2}-K)_+-u^*(s_1) -h^*(s_1)(s_2-s_1)  },
\eeaa

\no We have $\PP^* \in \widetilde{\cal M}$.
\end{lem}

\proof
The constraints are given by the Lagrange equations. For $s_1\in {\rm supp}(\P^1)$, the fact that $\P^*(S_1 = s_1) = \P^1[s_1]$ is given by the equation $\partial_{u(s_1)} \Gc(u^*,h^*,V^*)$. The fact that $\E^{\P^*}[S_2|S_1=s_1] = s_1$ is given by the equation $\partial_{h(s_1)} \Gc(u^*,h^*,V^*) = 0$. Finally, $\Cc^{K,{\rm bid}}\le \E^{\P^*}[(S_2-K)_+]\le \Cc^{K,{\rm ask}}$ is given by the equation $\partial_{V_K} \Gc(u^*,h^*,V^*) = 0$.
\ep

\begin{lem}\label{lemma:minG}
The map $\Gc$ reaches a minimum $\Gc^*$ at some $x^*\in \LL^1(\PP^1)^2\x\R^{|\Kc|}$ if and only if $\left(m_0, \P^1, \left(\Cc^{K,{\rm bid/ask}}\right)_{K\in\Kc}\right)$ is non-degenerate.
\end{lem}

\proof
Recall that as $u,h\in \LL^1(\PP^1)$ can be identified with vectors in $\R^{|{\rm supp}(\P^1)|}$ as $\P^1$ has finite support.

\no\underline{\rm Step 1:} We assume that $\left(m_0, \P^1, \left(\Cc^{K,{\rm bid/ask}}\right)_{K\in\Kc}\right)$ is non-degenerate. Let $\Cc\in \R^{|\Kc|}$ be a valid call prices vector. Let us prove that $\Gc$ reaches a minimum at some $x^*\in \R^{2|{\rm supp}(\P^1)|+|\Kc|}$. First we prove that $\lim_{|x|\to\infty}\Gc(x) = \infty$. Let $(x_n)_{n\ge 0}\subset \R^{2|{\rm supp}(\P^1)|+|\Kc|}$ such that $|x_n|\longrightarrow\infty$. We assume for contradiction that up to replacing $x_n$ by a subsequence, $\Gc(x_n)$ is bounded from above by $A>0$. Then up to taking a subsequence of $(x_n)$, we may assume that $\frac{x_n}{|x_n|}$ converges to some $x\in \Uc:=\big\{x'\in\R^{2|{\rm supp}(\P^1)|+|\Kc|}\big\}$. Now let the random vector
\b*
\Delta &:=& \big((\delta_{S_1 = s_1})_{s_1\in{\rm supp}(\P^1)},(\delta_{S_1 = s_1}(S_2-S_1))_{s_1\in{\rm supp}(\P^1)},((S_2-K)_+)_{K\in\Kc}\big),
\e*
so that for $\big((x_{1, s_1})_{s_1\in{\rm supp}(\P^1)},(x_{2, s_1})_{s_1\in{\rm supp}(\P^1)}, (x_{3, K})_{K\in\Kc}\big):= x \in\R^{2|\X_1|+|\Kc|}$, we have
\b*
x\cdot\Delta &=& x_{1, S_1}+x_{2, S_1}(S_2-S_1)+\sum_{K\in\Kc}x_{3, K}(S_2-K)_+.
\e*
and
\b*
\Gc(x) &=& \sum_{s_1\in{\rm supp}(\P^1)}x_{1,s_1}\P^1[s_1] +\sum_{K \in {\cal K}}f_2^{K,\mathrm{bid}/\mathrm{ask}}(x_{3, K},\omega_K)+\sum_{K \in {\cal K}}  x_{3,K} \Cc^{K,\mathrm{mid}}+\int e^{-x\cdot \Delta}d{m_0}.
\e*
Notice that as $\Cc^{K_i,{\rm bid}}\le \Cc_i\le \Cc^{K_i,{\rm ask}}$ for $1\le i \le k$, we have that $(\Cc_i)_{1\le i \le k}$ is the subgradient of $V\longmapsto \sum_{K \in {\cal K}}f_2^{K,\mathrm{bid}/\mathrm{ask}}(V_K,\omega_K)+\sum_{K \in {\cal K}}  V_K \Cc^{K,\mathrm{mid}}$ at some point $V^0\in\R^{|\Kc|}$. Then if we denote $b:=\sum_{K \in {\cal K}}f_2^{K,\mathrm{bid}/\mathrm{ask}}(V^0_K,\omega_K)+\sum_{K \in {\cal K}}  V^0_K \Cc^{K,\mathrm{mid}}$ and $a:= \big((\P^1[s_1])_{s_1\in{\rm supp}(\P^1)},0,(\Cc_i)_{1\le i \le k}\big)$, we have
\be\label{eq:minoration_G}
\Gc(x)\ge a\cdot x+b+\int e^{-x\cdot \Delta}d{m_0}.
\ee

\no\underline{\rm Case 1:} We may find $(s_1,s_2)\in {\rm supp}(\P^1)\x\R_+$ such that $x\cdot \Delta(s_1,s_2)<0$. As $x\cdot \Delta(s_1,\cdot)$ is affine by parts, we may find $\eps >0$ and an open interval $s_2\in I\subset \R$ such that $x\cdot \Delta(s_1,\cdot)\le -\eps$ on $I$. Then for $x'$ close enough to $x$, we have $x'\cdot \Delta(s_1,\cdot)\le -\frac12\eps$ on $I$. Then by \eqref{eq:minoration_G}, for $n$ large enough we have
\b*
\Gc(x_n)&\ge& a\cdot x_n+b+\int e^{-x_n\cdot \Delta}d{m_0}\\
&\ge&a\cdot x_n+b+{m_0}[\{y_1\}\x I]e^{|x_n|\frac12\eps}.
\e*
Therefore, by the fact that ${m_0}[\{s_1\}\x I]>0$, we have that $\Gc(x_n)$ diverges to $\infty$ as $|x_n|\longrightarrow\infty$, a contradiction.

\no\underline{\rm Case 2:} $x\cdot \Delta \ge 0$ on ${\rm supp}(\P^1)\x\R_+$. Then $\Gc(x_n)\ge a\cdot x_n +b = |x_n|a\cdot \frac{x_n}{|x_n|} +b$. As we assumed that $\Gc(x_n)$ is bounded and $\frac{x_n}{|x_n|}$ converges to $x$, we have
\be\label{eq:x_scalar}
a\cdot x \le 0.
\ee
We denote $(u,h,V):=x$, identifying $u$ and $h$ as functions ${\rm supp}(\P^1)\longrightarrow \R$, and we have
$$x\cdot \Delta = u(S_1)+h(S_1)(S_2-S_1)+\sum_{K\in\Kc}V_K(S_2-K)_+.$$
Let $\psi:=s\mapsto\sum_{K\in\Kc}V_K(s-K)_+$. We have $\psi(S_2)\ge -u(S_1)-h(S_1)(S_2-S_1)$. Then if we denote $f$, the convex hull of $\psi$ on $\R_+$, we have $\psi\ge f$ and for all $s_1\in\X_1$, we have $f(S_2)\ge -u(s_1)-h(s_1)\cdot(S_2-s_1)$. Therefore, $f\ge -u$ from last functional inequality computed in $S_2 = s_1$. By the fact that $f$ is the convex hull of $\psi$, which is piecewise affine, $f$ is also piecewise affine on the same intervals. Therefore, by using the notation $\Kc = (K_i)_{1\le i\le k}$ from Definition \ref{def:non-degen}, we may find $\lambda_i\ge 0$ for all $1\le i\le k$ such that $f = s\mapsto f(0)+\nabla f(0)s+\sum_{i=1}\lambda_i(s-K_i)_+$. Recall that by Definition \ref{def:non-degen} we have that $\Cc_0 = \E^{\P_1}[S_1]$, and $\Cc_{k+1} = 1$, thus we have
\b*
a\cdot x &=& \E^{\P^1}[u+f]+\sum_{i=0}^k \mu_i\Cc_i+\sum_{i=1}^{k+1}\lambda_i\big(\Cc_i-\E^{\P^1}[(S_2-K_i)_+]\big),
\e*
where $\mu_0 := -\nabla f(0)$, $\mu_{k+1}:= -f(0)$, and $\mu_i:=V_{K_i}-\lambda_i$ for $1\le i\le k$ are the unique coefficients such that
\be\label{eq:def-mu}
(\psi-f)(S_2) = \sum_{i=0}^k\mu_i(S_2-K_i)_+.
\ee
Notice that $(\psi-f)\ge 0$, therefore for all $0\le i\le k+1$, $\gamma_i := (\psi-f)(K_{i+1}) = (M_{\rm call}^t\mu)_i \ge 0$ by evaluating \eqref{eq:def-mu} in each $K_i$, where recall the definition of $M_{\rm call}$ and the conventions on $i=k+1$ from Definition \ref{def:non-degen}. Then $\mu = (M_{\rm call}^t)^{-1}\gamma$, and $\mu\cdot \Cc = (M_{\rm call}^{-1}\Cc)\cdot \gamma$. Finally,
$$a\cdot x = \E^{\P^1}[u+f]+(M_{\rm call}^{-1}\Cc)\cdot \gamma+\sum_{i=1}^k\lambda_i\big(\Cc_i-\E^{\P^1}[(S_2-K_i)_+]\big).$$
By \eqref{eq:x_scalar}, $a\cdot x$ is non-positive, therefore, the non-degeneracy of $\Cc$ gives that $\lambda_1 = ... = \lambda_k = 0$, $\gamma = 0$, and $u+f = 0$. Therefore $\mu = 0$, $f = 0$, $u = -f = 0$, $V = 0$. Therefore, $x\cdot \Delta = h(S_1)(S_2-S_1)$. By the fact that $x\cdot \Delta\ge 0$, $m_0-$a.e., (vi) from Definition \ref{def:non-degen} implies that we have that $h=0$. Finally $x\cdot\Delta = 0$ on $\X_1\x\R_+$, and finally $x=0$, which is a contradiction as $x\in \Uc$.

\no We proved that $\lim_{|x|\to\infty}\Gc(x) = \infty$. As $\Gc$ is convex, it reaches a minimum at some $x^*\in \R^{2|{\rm supp}(\P^1)|+|\Kc|}$.

\no\underline{\rm Step 2:} Now we assume that $\Gc$ reaches a minimum. Let us denote $x^*$ this minimum and let $\PP^*(ds_1,ds_2)={m_0}(ds_1,ds_2) e^{- \sum_{K \in {\cal K}} V_K^* (s_{2}-K)_+-u^*(s_1) -h^*(s_1)(s_2-s_1)  } $. By Lemma \ref{lemma:minGinM}, we have that $\P^*\in \widetilde{\cal M}$. Notice also that the measure $\P^*$ is equivalent to the measure $m_0$. Therefore, for all $i$ the map $\theta_i:= (S_2-K_i)_+- (S_1-K_i)_+-\mathbf{1}_{S_1\ge K_i}(S_2-S_1)$ is non-negative and non-(zero $\P^*-$a.e.). Therefore $\E^{\P^*}[\theta_i]>0$. Finally we observe that if we denote $\Cc_i:= \E^{\P^*}[(S_2-K)_+]$, then we have $\E^{\P^*}[\theta_i]=\Cc_i - \E^{\P^1}[(S_1-K)_+]$ from the martingale property of $\P^*$, and $\Cc^{K_i,{\rm bid}}\le \Cc_i\le \Cc^{K_i,{\rm ask}}$ as $\P^*\in \widetilde{\cal M}$.
Now for $1\le i\le k$, let $f_i$ the piecewise affine map such that $f$ is zero on $[0,K_{i-1}]$, with $f(K_i) =1$, affine on $[K_{i-1},K_i]$, $[K_{i},K_{i+1}]$, and $[K_{i+1},\infty)$, if $i\neq k$, and $f_k$ is constant equal to $1$ on $[K_k,\infty]$. We observe that for all $i$, $f_i$ is non-negative and non-zero $\P^*-$a.e. Furthermore, $0<\E^{\P^*}[f_i]=(M_{\rm call}^{-1}\Cc)_i$. We proved that $\left(m_0, \P^1, \left(\Cc^{K,{\rm bid/ask}}\right)_{K\in\Kc}\right)$ is non-degenerate as the other properties are obvious.
\ep

\no{\bf Proof of Theorem \ref{thm:minimG_elem_M}}
By introducing dual variables $u^{\rm bid},u^{\rm ask} \in (\RR_+)^{\Kc}$ for the inequalities for the call prices at bid and at ask, $\inf \Gc$ may be written as
\beaa
\inf \Gc&=&-\inf_{u^{\rm bid},u^{\rm ask} \in (\RR_+)^{\Kc}, v_K \in \RR,u,h \in \RR^{{\rm supp}(\P^1)}} \E^{\P^1}[u(S_1)]+\sum_{K \in {\cal K}_1} u_K^{\rm ask}\Cc^\mathrm{ask}_{K}-u_K^{\rm bid}\Cc^\mathrm{bid}_{K} +v_K\Cc^\mathrm{mid}_{K} \\
&&+ \frac{1 }{2}v_K^2 \omega_K+
\EE^{{m_0}}\left[  e^{-\sum_ {K \in {\cal K} } (u_K^{\rm ask}-u_K^{\rm bid} + v_K) (S_{1}-K)_+-u(S_1) -h(S_1) (S_{2}-S_1)}\right].
\eeaa
\no By setting  $v:=V-u^{\rm ask}+u^{\rm bid}$, the function, to be minimized, is equivalent to
\beaa
&&\E^{\P^1}[u(S_1)]+\sum_{K\in\Kc_1}u^{\rm ask}_K(\Cc_1^{K,\mathrm{ask}}-\Cc_1^{K,\mathrm{mid}})-u^{\rm bid}_K(\Cc_1^{K,\mathrm{bid}}-\Cc_1^{K,\mathrm{mid}})+V_K\Cc_1^{K,\mathrm{mid}}\\
&+&\frac{1 }{2}(V-u^{\rm ask}+u^{\rm bid})^2 \cdot\omega+
\EE^{{m_0}}\left[  e^{- \sum_{K \in {\cal K}}V_K (S_1-K)_+-u(S_1) -h(S_1) (S_{2}-S_1) }\right].
\eeaa
\no We observe that the minimization over $u^{\rm ask}$ and $u^{\rm bid}$ can be exactly performed and we obtain finally  an unconstrained optimization over $V$.

\no By Lemma \ref{lemma:minG}, the non degeneracy of $\left(m_0, \P^1, \left(\Cc^{K,{\rm bid/ask}}\right)_{K\in\Kc}\right)$ implies that $\Gc$ reaches a minimum. Let $\PP^* \in \widetilde{\cal M}$ from Lemma \ref{lemma:minGinM}, $\P^*$ is the optimiser of \eqref{opt1} from Proposition 1 in \cite{avellaneda1998minimum}.
\ep


\no{\bf Proof of Theorem \ref{thm:convergence}}
The first equivalence is given by Lemma \ref{lemma:minG}.

\no\underline{\rm Step 1:} The convergence result stems from an indirect application of Theorem 5.2 in \cite{beck2013convergence}. By a direct application of this theorem we get that
\be
&\Gc(x_k)-\Gc(x^*)\le \left(1-\frac{\sigma}{\min(L_1,L_2)}\right)^{n-1}\big(\Gc(x_0)-\Gc(x^*)\big),\label{eq:convergence_Beck2}
\ee
with $L_1$ (resp. $L_2$) is the Lipschitz constant of the $V-$gradient (resp. $(u,h)-$gradient) of $\Gc$, and $\sigma$ is the strong convexity parameter of $\Gc$. Furthermore, the strong convexity gives that
\be\label{eq:distance_from_conv}
|x_k-x^*|\le \sqrt{\frac{2}{\sigma}}\big(\Gc(x_k)-\Gc(x^*)\big)^{\frac12}.
\ee
Finally, by definition of $L_1$ and $L_2$, we have
\be\label{eq:control_gradient}
|\nabla \Gc(x_k)|\le (L_1+L_2)|x_k-x^*|.
\ee
These inequalities would prove the theorem with
\b*
\lambda = 1-\frac{\sigma}{\min(L_1,L_2)}&\mbox{and}&M=(L_1+L_2+1)\sqrt{\frac{2}{\sigma}}.
\e*
However the gradient $\nabla \Gc$ is locally but not globally Lipschitz, nor $\Gc$ strongly convex. Therefore we need to refine the theorem by looking carefully at where these constants are used in its proof.

\no\underline{\rm Step 2:} The constant $L_1$ is used for Lemma 5.1 in \cite{beck2013convergence}. We need for all $k\ge 0$ to have $\Gc(x_k)-\Gc(x_{k+1/2})\ge \frac{1}{2 L_1}|\nabla \Gc(x_k)|^2$. We want to find $C,L>0$ such that $\Gc(x_k)-\Gc\big(x_k-C\nabla\Gc(x_k)\big)\ge \frac{1}{2L}|\nabla\Gc(x_k)|$, then $L$ may be use to replace $L_1$ in the final step of the proof of Lemma 5.1 in \cite{beck2013convergence}. By the fact that $\lim_{|x|\to\infty}\Gc(x) = \infty$, the set ${\mathfrak C}(x_0):=\{x\in\R^{2|{\rm supp}(\P^1)|+|\Kc|}:\Gc(x)\le \Gc(x_0)\}$ is compact. Then $\nabla \Gc$ is bounded on ${\mathfrak C}(x_0)$. Therefore we may find $M_1>0$ such that for all $k$, we have $|\nabla\Gc(x_k)\cdot\Delta|\le M_1(1+|S_2|)$ where we denote $\Delta$ the random vector
\b*
\Delta &:=& \big((\delta_{S_1 = s_1})_{s_1\in{\rm supp}(\P^1)},(\delta_{S_1 = s_1}(S_2-S_1))_{s_1\in{\rm supp}(\P^1)},((S_2-K)_+)_{K\in\Kc}\big).
\e*
Furthermore, let $F(C):=\sup_{(u,x)\in\Uc\x{\mathfrak C}(x_0)}\int_0^1\int (u\cdot\Delta)e^{-x\cdot \Delta}e^{tCM_1(1+|S_2|)}dm_0dt$. We have
\b*
\Gc(x_k)-\Gc\big(x_k-C\nabla\Gc(x_k),h_k\big) &=& \nabla\Gc(x_k)\cdot \big(-C\nabla\Gc(x_k)\big)\\
&&-C^2\int_0^1 (1-t) D^2 \Gc\big(x_k-tC\nabla\Gc(x_k)\big)\big(\nabla\Gc(x_k)\big)^2 dt\\
&=&C|\nabla\Gc(x_k)|^2\\
&&-C^2\int_0^1 (1-t) \int (\nabla\Gc(x_k)\cdot\Delta)^2e^{-x_k\cdot\Delta} e^{tC\nabla\Gc(x_k)\cdot\Delta}dm_0 dt\\
&\ge& C|\nabla\Gc(x_k)|^2-C^2|\nabla\Gc(x_k)|^2F(C)\\
&=&\big(C-C^2F(C)\big)|\nabla\Gc(x_k)|^2.
\e*
As $F$ is non-decreasing finite, then when $C\longrightarrow 0$ we have $\frac{C-C^2F(C)}{C}\longrightarrow 1$. Then for $C$ small enough, let $L:= \frac{1}{2\big(C-C^2F(C)\big)}>0$. We get
\b*
\Gc(x_k)-\Gc\big(x_k-C\nabla\Gc(x_k)\big)\ge \frac{1}{2L}|\nabla\Gc(x_k)|^2.
\e*

\no\underline{\rm Step 3:} The constant $\sigma$ is used to get the result from (3.21) in \cite{beck2013convergence}. Then we just need the inequality
\be\label{eq:ineq_sec_order}
\Gc(y)\ge \Gc(x)+\nabla \Gc(x)\cdot(y-x)+\frac{\sigma}{2}|y-x|^2,
\ee
to hold for some $y =x^*$ and $x = x_k$ for all $k\ge 0$. Now we give a lower bound for $\sigma$. The map $(u,x)\longmapsto D^2\Gc(x)u^2=\int (u\cdot\Delta)^2e^{-x\cdot\Delta}dm_0>0$ is continuous on $\Uc\x {\mathfrak C}(x_0)$ compact, therefore it has a lower bound $\sigma>0$. This constant also works for \eqref{eq:distance_from_conv}. Similar, $\sup_{(u,x)\in \Uc\x {\mathfrak C}(x_0)}D^2\Gc(x)u^2$ may replace $L_1+L_2$ from \eqref{eq:control_gradient}.

\no\underline{\rm Step 4:} Finally, as we focus on the $L_1$ optimization phase, we may replace $n-1$ by $n$ in the convergence formula \eqref{eq:convergence_Beck2}, see the proof of Theorem 5.2 in \cite{beck2013convergence}.

\no Now the existence of $M>0$ stems from the facts that $\Gc(x_k)-\Gc(x^*)\ge \frac12\sigma|x_k-x^*|^2$, and $|\nabla\Gc(x_k)|\le L|x_k-x^*|$.

\no\underline{\rm Step 5:} Now we just use the fact that
\beaa 
\nabla \Gc &=& \sum_{s_1\in{\rm supp}(\P^1)}\left(\P^1[\{s_1\}]-\P_n\circ (S_1)^{-1}[\{s_1\}]\right)e_{s_1}\\
&&+\sum_{s_1\in{\rm supp}(\P^1)}\left(\E^{\P_n}[S_2-s_1,S_1=s_1]\right)e_{|{\rm supp}(\P^1)|+s_1}\\
&&+\sum_{i=1}^k\left(C_K-\E^{\P_n}[(S_2-K_i)_+]\right)e_{2|{\rm supp}(\P^1)|+i},
\eeaa 
where $C_K\in[\Cc^{K,{\rm bid}},\Cc^{K,{\rm ask}}]$, and $(e_i)_{1\le i \le 2|{\rm supp}(\P^1)|+|\Kc|}$ is the canonical basis. Therefore, this gradient is a crucial estimate of the mismatch of $\P$ in terms of first marginal, martingale property, and correctness of the call prices it gives. However, as we consider $x_n$ after the Bregman projection from the Sinkorn's algorithm on $u_n$ and $h_n$, $\P_n$ is martingale as minimal in $h_n$, and has marginal $\P^1$ on $S_1$ as minimal in $u_n$. Then $\nabla \Gc =\sum_{i=1}^k\left(C_K-\E^{\P_n}[(S_2-K_i)_+]\right)e_{2|{\rm supp}(\P^1)|+i}$. We obtain the inequality by taking the infinite norm on this vector, that is equivalent to any other as we are in finite dimensions. Therefore up to raising $M>0$, the result is proved.
\ep

\no{\bf Proof of Theorem \ref{thm:minimG_elem_M_all}}
We build the probabilities $\P_i$ by induction. We first set $\P_0:= \delta_{S_0}$.

\no Now let $1\le i\le n-1$ we assume that $\P_0,...,\P_{i-1}$ are created, that the results of the Theorem is proved for them, and that they satisfy that $\left(m^{j-1,j}, \P_{j-1}, \left(\Cc^{K,{\rm bid/ask}}\right)_{K\in\Kc_{j}}\right)$ is non-degenerate for $j\le i-1$.

\no We first prove the non-degeneracy of $\left(m^{i-1,i}, \P_{i-1}, \left(\Cc^{K,{\rm bid/ask}}\right)_{K\in\Kc_{i}}\right)$. For this we need to prove \rm{(i)} to \rm{(vi)} from Definition \ref{def:non-degen}. Thanks to the non-degeneracy of $\left(m_n, \left(\Cc^{K,{\rm bid/ask}}_i\right)_{1\le i\le n,K\in\Kc_i}\right)$, we set $\Cc:=\Cc^{i}$ from Definition \ref{def:non-degen-all}.

\no\underline{\rm (i)} holds by {\rm (i)} of Definition \ref{def:non-degen-all} for $1 \le l\le k$. For $l=0$, if $i=1$, the result is trivial as $\E^{\P_0}[S_0] = S_0$. For $i\ge 1$, we have that $\E^{\P_{i-2}}[S_{t_{i-1}}] = S_0$ by induction assumption, then as $\P_{i-1} = \P_{i-2}\P^{i-2,i-1}(ds_{i-1}|s_{i-2})$ with $\P^{i-2,i-1}$ martingale by induction assumption together with Definition \ref{def:non-degen-all}, we have
\beaa 
\Cc^{0,{\rm bid}}_i = \Cc^{0,{\rm ask}}_i = \E^{\P_{i-1}}[S_{t_{i-1}}] = S_0  = \E^{\P_{i-2}}[S_{t_{i-1}}] = S_0.
\eeaa 
The case $l=k+1$ is also trivial from Definition \ref{def:non-degen-all}.

\no\underline{\rm (ii)} holds by {\rm (ii)} of Definition \ref{def:non-degen-all}.

\no\underline{\rm (iii)} Let $1\le l\le k$, we have that $(S_{t_{i-1}}-K_l)_+\le (S_{t_{i-1}}-K^{i-1}_{l'})_++(K^{i-1}_{l'}-K_{l})_+$ where we take $K^{i-1}_{l'}$ from \rm{(iii)} in Definition \ref{def:non-degen-all}. Then we have thanks to \rm{(iii)} in Definition \ref{def:non-degen-all}:
\beaa 
\E^{\P_{i-1}}[(S_{t_{i-1}}-K_l)_+] &\le& \E^{\P_{i-1}}[(S_{t_{i-1}}-K^{i-1}_{l'})_+]+(K^{i-1}_{l'}-K_{l})_+\\
&\le & \Cc^{K^{i-1}_{l'},\rm ask}_{i-1}+(K^{i-1}_{l'}-K_{l})_+\\
&< & \Cc^{K_l,\rm bid}\\
&\le & \Cc_l.
\eeaa 
\rm (iii) is proved.

\no\underline{\rm (iv)} As by induction assumption, $\frac{d\P^{i-2,i-1}}{dm^{i-2,i-1}}>0$, $dm^{i-2,i-1}-$a.s. Therefore,
\beaa 
{\rm supp \P^{i-1}}&=&{\rm supp \P^{i-2,i-1}\circ S_{t_{i-1}}^{-1}}\\
&=& {\rm supp m^{i-2,i-1}\circ S_{t_{i-1}}^{-1}}\\
&=&{\rm supp m^{i-1,i-2}\circ S_{t_{i-1}}^{-1}}.
\eeaa 
by \rm{(iv)} of Definition \ref{def:non-degen-all}. \rm{(iv)} is proved.

\no\underline{\rm (v)} holds by {\rm (v)} of Definition \ref{def:non-degen-all}.

\no\underline{\rm (vi)} holds by {\rm (vi)} of Definition \ref{def:non-degen-all}.

The non degeneracy of $\left(m^{i-1,i}, \P_{i-1}, \left(\Cc^{K,{\rm bid/ask}}\right)_{K\in\Kc_{i}}\right)$ allows to apply Theorem \ref{thm:minimG_elem_M}, which gives the rest of the result of the Theorem for $i-1$. The Theorem is then proved by induction.
\ep

\newpage
\bibliographystyle{plain}
\bibliography{mabib}
\end{document}